\documentclass{jpp}

\usepackage{graphicx}
\graphicspath{{./}{figures/}}
\usepackage{amsmath}
\usepackage{mathtools}
\usepackage{physics}
\usepackage{bm}
\usepackage{booktabs}
\usepackage{multirow}
\usepackage{diagbox}
\usepackage[position=top]{subfig}
\usepackage{array}

\usepackage{tikz}
\usepackage{tcolorbox}
\usepackage{xcolor}
\usepackage{enumitem}
\usepackage{footmisc}
\usepackage{appendix}
\usepackage{mwe}
\usepackage{natbib}

\usepackage{hyperref}
\hypersetup{
    colorlinks=true,
    linkcolor=blue,
    citecolor=blue,
    urlcolor=blue,
    pdftitle={Your Document Title},
    pdfauthor={Your Name}
}

\providecommand{\url}[1]{\href{#1}{#1}}
\providecommand{\dodoi}[1]{doi:~\href{http://doi.org/#1}{\nolinkurl{#1}}}
\providecommand{\doeprint}[1]{\href{http://ascl.net/#1}{\nolinkurl{http://ascl.net/#1}}}
\providecommand{\doarXiv}[1]{\href{https://arxiv.org/abs/#1}{\nolinkurl{https://arxiv.org/abs/#1}}}

\definecolor{darkgreen}{rgb}{0.0, 0.5, 0.0}

\newcommand{\code}[1]{\texttt{#1}}

\begin{document}

\title{Particle injection in three-dimensional relativistic magnetic reconnection\footnote{Released \today}}

\author{Omar French}
% \homepage{National Science Foundation Graduate Research Fellow.}
\author{Omar French\aff{1}
\corresp{\email{omar.french@colorado.edu}},
Gregory R. Werner\aff{1}
 \and Dmitri A. Uzdensky\aff{2}
}

\affiliation{\aff{1}Center for Integrated Plasma Studies, Department of Physics, 390 UCB, University of Colorado, Boulder, CO 80309-0390, USA
\aff{2}Rudolf Peierls Centre for Theoretical Physics, University of Oxford, Oxford OX1 3NP, UK}

\maketitle

\begin{abstract}
Relativistic magnetic reconnection has been proposed as an important nonthermal particle acceleration (NTPA) mechanism that generates power-law spectra and high-energy emissions. Power-law particle spectra are in general characterized by three parameters: the power-law index, the high-energy cutoff, and the low-energy cutoff (i.e., the injection energy). Particle injection into the nonthermal power law, despite also being a critical step in the NTPA chain, has received considerably less attention than the subsequent acceleration to high energies. Open questions on particle injection that are important for both physical understanding and astronomical observations include how the upstream magnetization~$\sigma$ influences the injection energy and the contributions of the known injection mechanisms (i.e., direct acceleration by the reconnection electric field, Fermi kicks, and pickup acceleration) to the injected particle population. Using fully kinetic particle-in-cell simulations, we uncover these relationships by systematically measuring the injection energy and calculating the contributions of each acceleration mechanism to the total injected particle population. We also present a theoretical model to explain these results. Additionally, we compare two- and three-dimensional simulations to assess the impact of the flux-rope kink and drift-kink instability on particle injection. We conclude with comparisons with previous work and outlook for future work. 
\end{abstract}
\keywords{astrophysical plasmas --- plasma simulation}

\section{Introduction} \label{sec:intro}

The origin of high-energy emissions in the Universe is often a nonthermal power-law spectrum of relativistic particles. Therefore, to explain high-energy emissions, it is necessary to understand how power-law distributions of nonthermal particles are populated and sustained, with the former being the concern of injection studies and the latter being the concern of studies about nonthermal particle acceleration to high energies. 

The question of how a power-law distribution of particles is populated may be decomposed into two questions: 
(1) Under what conditions are particles injected, i.e., eligible to participate in a continual, power-law-forming acceleration process? (2) What physical mechanisms are responsible for injection?

Regarding the first question, the injection criterion has typically been expressed since~\citet{Fermi1949} as an energy threshold~$\gamma_{\rm inj}$ that a particle must surpass, so that particles with energy~$\gamma > \gamma_{\rm inj}$ are ``injected" and can experience further Fermi acceleration to even higher energies, whereas those with~$\gamma < \gamma_{\rm inj}$ cannot. Physically, this may be owed to the relativistic gyroradius of particles scaling linearly with~$\gamma$, which facilitates acceleration via stochastic scattering off of turbulent fluctuations~\citep{Lemoine_2020}. In this study, we also presuppose that~$\gamma_{\rm inj}$ exists and define it as the low-energy boundary of the nonthermal power-law segment of the downstream particle distribution function. Thus, the first question is recast as a problem of relating~$\gamma_{\rm inj}$ to the system parameters and explaining the connection. 

Accordingly, the second question is also recast: it is to understand what ``injection mechanisms"---mechanisms that accelerate particles from a thermal upstream to or beyond the injection energy~$\gamma_{\rm inj}$ that gates the nonthermal power-law spectrum---are active, how much work they do to each particle, and upon how many particles they act.

These questions of particle injection have been largely unresolved across several processes that are thought to power high-energy emissions in relativistic astrophysical environments, such as jets from active galactic nuclei, 
pulsar wind nebulae, 
neutron star magnetospheres, 
and accreting black hole coronae. 
These processes include relativistic turbulence \citep{Chandran_2000,Zhdankin_2017,Zhdankin_2018,Zhdankin_2019,Comisso2018,Comisso2019,Lemoine2019,Wong2020,Demidem_2020,Lemoine_2020,Guo2025_review,Mehlhaff_2025}, collisionless relativistic shocks \citep{Blandford_1978,Blandford_1987,Spitkovsky_2008,Sironi2011,Caprioli_2014,Parsons_2023}, and relativistic magnetic reconnection \citep{Bulanov1976,Blackman_1994,Zenitani2001,Jaroschek2004b,Giannios2010,Cerutti2013,Sironi2014,Melzani2014,Guo2014,Guo2015,Guo2016,Guo2019,Nalewajko2015,Werner2016,Werner2017,Werner2018,Rowan2017,Petropoulou2018,Schoeffler2019,Mehlhaff2020,Kilian2020,Hakobyan2021,Werner2021,Sironi2022,Uzdensky2022,Li2023,French_2023,HaoZhang_2023,Gupta_2025} \citep[see][for recent reviews]{Guo_2023,Sironi_2025}. 

In the context of relativistic magnetic reconnection, several injection and high-energy acceleration mechanisms have been studied, both analytically and numerically via fully kinetic particle-in-cell (PIC) simulations: ``direct" acceleration by the parallel electric field with a finite guide magnetic field (i.e., a finite non-reversing, out-of-plane component of the magnetic field) near X-points \citep{Larrabee2003,Zenitani2005,Zenitani2008,Cerutti2013,Cerutti2014,Ball2019,Sironi2022,Totorica_2023,Gupta_2025}, Speiser orbits in the case of zero guide field \citep{Speiser1965,Hoshino2001,Zenitani2001,Uzdensky2011b,Cerutti2012,Cerutti2013,Cerutti2014,Nalewajko2015,Uzdensky2022}, Fermi acceleration \citep{Fermi1949,Drake2006,Giannios2010,Guo2014,Guo2015,Dahlin2014,Qile_2021,French_2023,Qile_2024}, parallel electric field acceleration in the exhaust region \citep{Egedal2013,Zhang2019}, and acceleration by the pickup process in the exhaust region \citep{Drake2009,Sironi2020,French_2023,Chernoglazov2023}. Previous work on particle injection in magnetic reconnection has included studies in the transrelativistic regime for a proton-electron plasma with a weak guide field \citep{Ball2019,Kilian2020} and relativistic pair plasmas for various guide-field strengths \citep{Sironi2022,Guo2022_comment,French_2023,Totorica_2023} and upstream magnetizations \citep{Sironi2022,Guo2022_comment,Totorica_2023,Gupta_2025}.

However, these studies have not elucidated several important aspects of injection, such as how the injection stage in relativistic magnetic reconnection is influenced by the upstream magnetization and 3D effects. Addressing this key question is the subject of this paper. We achieve this using analytical theory and fully kinetic 2D and 3D PIC simulations of relativistic magnetic reconnection in nonradiative collisionless pair plasmas.
To uncover spectral quantities, most importantly the injection energy~$\gamma_{\rm inj}$, we run a spectral fitting procedure that is improved from our previous work \citep{French_2023}. To uncover the relative contributions of different injection mechanisms, we employ a methodology similar to \citet{French_2023}, wherein we consider several mechanisms, namely, parallel electric fields near X-points, Fermi kicks by the motional electric field, and the pickup process. 

Throughout this paper, we use the units~$m_e = c = 1$. That is, we normalize velocities to the speed of light~$c$, momenta to~$m_e c$, and energies to the electron rest energy~$m_e c^2$. Furthermore, we denote~$\gamma$ as the energy of a single particle, $E$ as the sum or integral of energies of multiple particles, and $\textbf{v}$ as the particle 3-velocity in the simulation frame. Primed vector quantities, such as the velocity~$\textbf{v}'$ or the momentum~$\textbf{p}' \equiv \gamma' \textbf{v}'$, denote a Lorentz boost to the~$\textbf{v}_E \equiv \textbf{E} \times \textbf{B}/B^2$ drift-velocity frame, where~$\textbf{E}$ and~$\textbf{B}$ respectively denote the electric and magnetic field vectors. This is done to eliminate the context of bulk plasma motions. Primed scalar quantities denote those whose vector inputs are primed, e.g.~$\gamma' \equiv (1 - \textbf{v}' \cdot \textbf{v}')^{-1/2}$. Subscripts~$\parallel, \perp$ indicate components relative to the local magnetic field in the laboratory frame ($E \times B$ drift-velocity frame) if the quantity is not primed (primed). Lastly, we reference four dimensionless quantities used throughout this paper: the cold upstream magnetization
\begin{equation} \label{eq:sigma}
\sigma \equiv B_0^2/4\pi nm_ec^2, 
\end{equation}
corresponding to the average magnetic energy per particle rest mass energy (where~$B_0$ is the ambient upstream magnetic field and ~$n \equiv n_i + n_e$ is the total upstream plasma density), 
the hot upstream magnetization
\begin{equation} \label{eq:sigmah}
\sigma_h \equiv B_0^2/4\pi h, 
\end{equation}
where $h$ is the relativistic plasma enthalpy density, the ambient upstream temperature normalized by particle rest mass energy
\begin{equation}
\theta_0 \equiv k_BT_0/m_e c^2,    
\end{equation}
and the normalized guide-field strength
\begin{equation}
    b_g \equiv B_g/B_0,
\end{equation}
where~$B_g$ is the ``guide magnetic field," a uniform, non-reversing component of the magnetic field that orients out-of-plane.

The rest of this paper is organized as follows. Section~\ref{sec:injection_picture} presents our theoretical picture of particle injection. Section~\ref{sec:setup} describes the setup for the simulations. Section~\ref{sec:results} contains the analysis and results from each simulation. Section~\ref{sec:discussion} discusses comparisons with previous work and outlooks for future work. Section~\ref{sec:conclusions} presents our main conclusions.
\section{Theoretical picture of particle injection} \label{sec:injection_picture}

In concordance with our overview in Section~\ref{sec:intro}, our picture of particle injection in relativistic magnetic reconnection has two components. The first is an analytical model for the injection energy~$\gamma_{\rm inj}$ (Section~\ref{ss:injection_energy_model}). The second is a detailed description of each mechanism that injects particles (Section~\ref{ss:injection_mechanisms}).

\subsection{Injection criterion} \label{ss:injection_energy_model}

Suppose the criterion for a particle to be ``injected", regardless of the mechanism of injection, is that its gyroradius~$r_g$ exceeds the reconnection layer half-thickness at the X-point~$\delta/2$, i.e., $r_g \geq \delta/2 \implies \gamma \geq \gamma_{\rm inj}$ \citep{Speiser1965,Giannios2010}. 
This criterion ensures that a gyrating particle centered on the X-point spends some fraction of each orbit outside the layer, which causes it to adopt a meandering Speiser trajectory. It follows that a newly injected particle will satisfy
\begin{equation} \label{eq:r_g}
    \delta = 2 r_g(\gamma = \gamma_{\rm inj}) = 2 \gamma_{\rm inj} v_\perp \frac{m_e c}{eB} = 
    2 \gamma_{\rm inj} v_\perp \, \omega_{\rm ce}^{-1} \big( 1 + b_g^2 \big)^{-1/2} \simeq 
    2 \gamma_{\rm inj} c \, \omega_{\rm ce}^{-1} \big( 1 + b_g^2 \big)^{-1/2}
\end{equation}
\begin{equation} \label{eq:ginj_refactor}
    \implies \gamma_{\rm inj} \simeq \frac{\delta}{2} \frac{\omega_{\rm ce}}{c} \sqrt{1 + b_g^2} = \frac{1}{2} \frac{\delta}{\rho_0} \sqrt{1 + b_g^2},
\end{equation}
where~$\omega_{\rm ce} \equiv eB_0/m_e c$ is the nominal electron gyrofrequency defined with the upstream magnetic field~$B_0$ and~$\rho_0 \equiv c/\omega_{\rm ce} = m_e c^2/eB_0$ is the corresponding nominal relativistic gyroradius. In Eq.~\eqref{eq:r_g} we have ignored pitch angle corrections and assumed~$v_\perp \simeq c$. 

Our next task is to estimate the layer thickness~$\delta$ at the X-point. We shall take~$\delta \simeq d_{\rm e, rel}$, where~$d_{\rm e, rel}$ is the relativistic collisionless electron skin depth inside the layer, and assume that the electron density in the layer is comparable to the ambient upstream electron density~$n_e$. Additionally, we assume that the velocity distribution is isotropic in the layer. Then,
\begin{equation} \label{eq:d_e}
    \delta \simeq d_{\rm e, rel} = c \, \omega_{\rm pe, rel}^{-1}
    = \sqrt{\Gamma} \, c \, \omega_{\rm pe}^{-1},
\end{equation}
where~$\omega_{\rm pe} \equiv \big[ 4\pi n_e e^2 / m_e \big]^{1/2}$ is the corresponding electron plasma frequency and~$\Gamma$ is the mean Lorentz factor of electrons in the layer. 

Substituting Eq.~\eqref{eq:d_e} into Eq. \eqref{eq:ginj_refactor}, we obtain
\begin{equation} \label{eqn:ginj_model_proper}
    \gamma_{\rm inj} \simeq \sqrt{\Gamma \, \sigma \, (1 + b_g^2)/2},
\end{equation}
where for a pair plasma~$\sigma = \sigma_e/2 = \omega_{\rm ce}^2/(2\omega_{\rm pe}^2)$ is the upstream magnetization.

The remaining task is to model~$\Gamma$, which can be represented as the sum of two contributions 
\begin{equation} \label{eq:gamma}
    \Gamma = \Gamma_{\rm up} + \frac{\langle W \rangle}{m_e c^2},
\end{equation}
where~$\Gamma_{\rm up}$ is the average Lorentz factor of particles arriving from the upstream and~$\langle W \rangle$ is the average work done to particles while they are in the layer. If the upstream particle distribution is thermal, then~$\Gamma_{\rm up} = \Gamma_{\rm th} = K_3(1/\theta_0)/K_2(1/\theta_0)$, where~$\Gamma_{\rm th} = K_3(1/\theta_0)/K_2(1/\theta_0)$ is the ``thermal Lorentz factor", i.e., the average energy of electrons in a Maxwell-J\"uttner distribution of upstream temperature~$\theta_0 \equiv k_B T_0/m_e c^2$.

Meanwhile, $\langle W \rangle$ can generally be written~$\langle W \rangle = k\sigma m_ec^2$, where one expects the dimensionless coefficient~$k>0$ to be a constant independent of~$\sigma$. We may constrain~$k$ as follows. First, $U_B = B_0^2/8\pi$ implies that the available magnetic energy per particle is~$U_B/n = B_0^2/(8\pi n) = (\sigma/2) m_e c^2$. Second, assuming that the efficiency of energy conversion is~$\sim 50\%$, the average dissipated magnetic energy per particle is~$\simeq (\sigma/4) m_ec^2$. Third, since the particles are energized partially as they enter the current sheet and partially as they exit the current sheet (e.g., by magnetic tension release in the outwards-accelerating plasma), we may assume that the average dissipated magnetic energy per particle \textit{that exists in the current sheet} is~$\lesssim (\sigma/8) m_ec^2$. This gives a constraint of~$k \lesssim 1/8$. 
We stress that, while the exact value of~$k$ is uncertain, the above argument allows us to treat~$k$ as a small parameter. Accordingly, we shall retain~$k$ in the rest of this subsection.

We can now write Eq.~\eqref{eq:gamma} as
\begin{equation} \label{eq:Gamma}
    \Gamma = \Gamma_{\rm th} + k \, \sigma \simeq \Gamma_{\rm th} \big[ 1 + k \, \sigma_h \big] = \sigma \big[ \sigma_h^{-1} + k \big], 
\end{equation}
where we approximate~$\sigma_h \simeq \sigma/\Gamma_{\rm th}$. We see that there are two distinct relativistic ($\sigma_h \gg 1$) regimes based on which of the two terms in Eq.~\eqref{eq:Gamma} dominates, i.e., how~$\sigma_h$ compares to~$k^{-1}$: 
\medskip
\noindent
\begin{enumerate}[label=(\roman*), leftmargin=30pt, itemindent=0pt, labelsep=10pt]
    \item \textbf{Thermally-dominated regime~$1 \lesssim \sigma_h \lesssim k^{-1}$}. 
    In this moderate-magnetization case (covering~$\gtrsim 1$ decade in~$\sigma_h$), the average particle energy in layer is~$\Gamma \simeq \Gamma_{\rm th}$, i.e., inherited from the thermal upstream and not controlled by~$\sigma$ or reconnection. In this regime, the upstream plasma conditions fully govern the layer thickness and partially govern the injection energy~$\gamma_{\rm inj}$. 
    In particular, the injection energy according to Eq.~\eqref{eqn:ginj_model_proper} becomes 
    \begin{equation} \label{eqn:ginj_model_soft}
        \gamma_{\rm inj} \simeq \sigma \sqrt{\sigma_h^{-1}  \, (1 + b_g^2)/2}.
    \end{equation}
    In particular, 
    \begin{equation}
    \gamma_{\rm inj} \simeq
    \begin{cases}
    \sqrt{\sigma(1 + b_g^2)/2}, & \theta \ll 1 \\
    \sqrt{\Gamma_{\rm th} \sigma (1 + b_g^2)/2}, & \theta \gg 1
    \end{cases}
    \end{equation}

    \item \textbf{Magnetically-dominated regime~$\sigma_h \gg k^{-1}$.} 
    In this case, the average particle energy in the layer~$\Gamma$ is dominated by~$\langle W \rangle/m_e c^2$ and upstream contributions can be ignored. Hence Eq.~\eqref{eqn:ginj_model_proper} becomes
    \begin{equation} \label{eqn:ginj_model_hard}
        \gamma_{\rm inj} \simeq \sigma \sqrt{k (1 + b_g^2)/2}.
    \end{equation}
\end{enumerate}

There are several implications for the resulting dynamic range of the power-law spectrum. Assuming that~$\gamma_c = C \sigma$ \citep{Werner2016}, one expects
\begin{equation}
R \equiv \gamma_c/\gamma_{\rm inj} \simeq 
\begin{cases}
\sqrt{2}C \, \sigma_h^{1/2} \left[ 1 + b_g^2 \right]^{-1/2} & 1 \lesssim \sigma_h \lesssim k^{-1} \\
\sqrt{2}C \, k^{-1/2} \left[ 1 + b_g^2 \right]^{-1/2} & \sigma_h \gg k^{-1} \\
\end{cases}
\end{equation}
 
In this model, the explicit guide-field dependence of~$\gamma_{\rm inj}(b_g) \propto \sqrt{1 + b_g^2}$ is owed to stronger magnetic fields decreasing the particle gyroradius, resulting in a greater energy necessary to satisfy~$r_g \geq \delta/2$. 
There could also be an implicit guide-field dependence if the layer thickness depends on guide-field strength. 
This is plausible because a strong guide field can prolong the duration over which particles remain in the X-point, causing~$k$ to increase. Nevertheless, as we shall find in Section~\ref{ss:ginj_w_sigma}, the scaling~$\gamma_{\rm inj}(b_g) \propto \sqrt{1 + b_g^2}$ is in rough agreement with \citet{French_2023}.

\subsection{Injection mechanisms} \label{ss:injection_mechanisms}

\begin{figure}
    \centering
    \includegraphics[width=\textwidth]{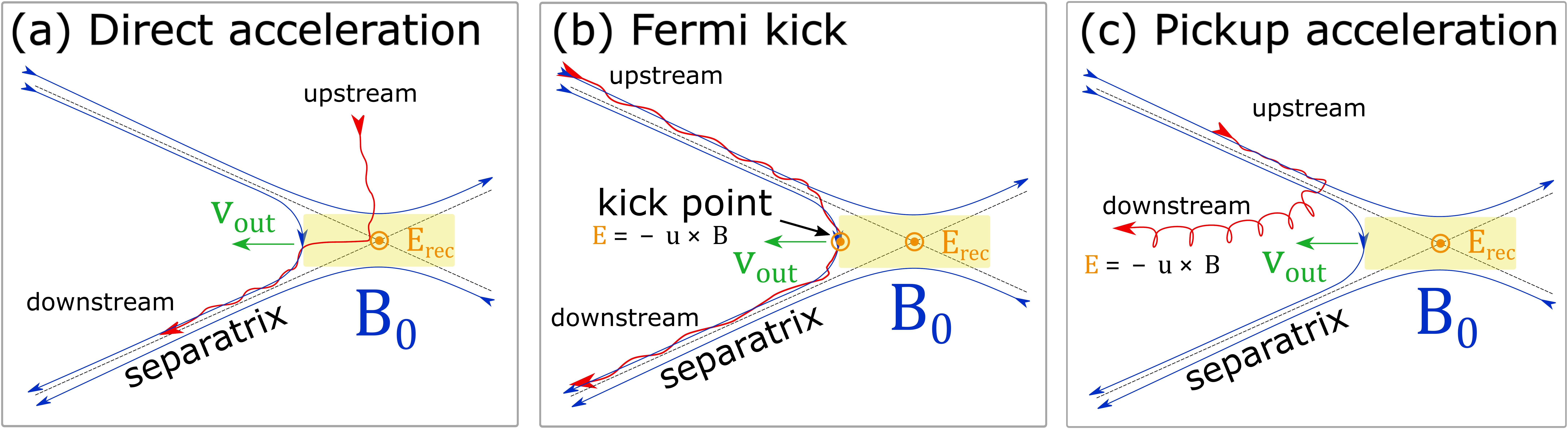}
    \caption{Cartoons of several particle injection mechanisms, adapted from~\cite{French_2023}. In each panel, $B_0$ is the reconnecting magnetic field, $E_{\rm rec}$ is the reconnection electric field, and~$v_{\rm out}$ is the reconnection outflow speed. 
    (a) Injection by direct acceleration from the reconnection electric field near an X-point. (b) Injection by a Fermi ``kick." (c) Injection by the pickup process, wherein~$\lvert \textbf{p}_\perp'\rvert$ suddenly increases upon crossing the separatrix and subsequent entry into the downstream region.}
    \label{fig:inj_mech_cartoon}
\end{figure}

In our first paper \citep{French_2023}, we stipulated that particles can be injected only as they transition from upstream to downstream, i.e., as they cross or interact with the magnetic separatrix between these two regions. Applying this assumption in the context of a reconnection layer has led to the identification of the following three injection mechanisms \citep{French_2023}, illustrated in Figure~\ref{fig:inj_mech_cartoon}.

\medskip
\noindent
\begin{enumerate}[label=(\roman*), leftmargin=30pt, itemindent=0pt, labelsep=10pt]
    \item [(a)] \textbf{Direct acceleration}: Incoming particles are accelerated directly by the reconnection electric field~$E_{\rm rec}$ in microscopic diffusion regions around magnetic X-points (Figure~\ref{fig:inj_mech_cartoon}a).     When immersed in~$E_{\rm rec}$, charged particles gain energy at a rate of~$\dot{W}_{\rm direct} = \big( \eta_{\rm rec} \beta_A \omega_{\rm ce}\big) m_e c^2$, 
    where~$\eta_{\rm rec} \beta_A \equiv E_{\rm rec}/B_0 \simeq 0.1$ (when the guide magnetic field is weak) is the reconnection rate normalized by 
    the speed of light, where~$\beta_A \equiv \big[ \sigma_h/(1+\sigma_h) \big]^{1/2} \simeq \big[ \sigma/(1+\sigma) \big]^{1/2}$ is the dimensionless upstream Alfv\'en speed (provided the upstream plasma is relativistically cold, i.e. $\theta_0 \equiv k_B T_0 / m_e c^2 \ll 1$) and~$\omega_{\rm ce} \equiv eB_0/m_ec$. Hence
    \begin{equation} \label{eq:W_direct}
        W_{\rm direct}/m_ec^2 = \eta_{\rm rec} \beta_A \big(  \omega_{\rm ce} \Delta t\big) \simeq 0.1 \, \omega_{\rm ce} \Delta t.
    \end{equation}
    If a guide field~$b_g \equiv B_g/B_0 \gtrsim \eta_{\rm rec} \simeq 0.1$ is present, then~$E_{\rm rec}$ is well-approximated by the parallel electric field~$\lvert \textbf{E}_\parallel \rvert \equiv (\textbf{E} \cdot \textbf{B}) /\lvert \textbf{B} \rvert$ at the X-point where the in-plane magnetic field vanishes. In this case, particles gain significant parallel momentum~$\textbf{p}'_\parallel \equiv \gamma' \textbf{v}'_\parallel = \gamma \textbf{v}_\parallel = \gamma (\textbf{v} \cdot \textbf{B}) \textbf{B}/\lvert \textbf{B} \rvert^2$. 
    
    \item [(b)] \textbf{Fermi kick}: The relaxation of freshly reconnected magnetic field-line tension gives rise to a Fermi acceleration process wherein the local curvature drift velocity~$\textbf{v}_C$ of particles is aligned with the motional ideal-MHD electric field~$\textbf{E}_m = - \textbf{u} \times \textbf{B}$ associated with the rapid advection of the newly-reconnected magnetic field lines \citep{Drake2006,Dahlin2014,Guo2014,Li2018b,Kilian2020,Qile_2021,Majeski_2023,French_2023,Qile_2024}. This mechanism is illustrated in Figure~\ref{fig:inj_mech_cartoon}b.

    Consequently, an incoming particle experiences one Fermi reflection (or ``kick", i.e., one half-cycle of a continual Fermi process) that flips its direction in the $E \times B$ drift frame. If the Fermi kick is treated as a 1D collision, then the net energy gain is obtained by considering the particle's initial velocity in the outflow (i.e., $E \times B$-drift) frame projected onto the outflow direction, $\beta_i'$, negating it (i.e., $\beta_f' = -\beta_i'$), and boosting back to the simulation frame. Assuming that the outflow speed is the in-plane Alfv\'en speed, the result is a velocity gain of
    \begin{equation}
        \Delta \beta \equiv \beta_f - \beta_i = \frac{2\beta_{\rm Ax} \big[ 1 + \beta_i^2 \big] - 2\beta_i \big[ 1 + \beta_{\rm Ax}^2 \big]}{ \big[ 1 + \beta_{\rm Ax}^2 \big] -2 \beta_i \beta_{\rm Ax} },
    \end{equation}
    where~$\beta_j \equiv v_j/c$ for any subscript~$j$ and~$\beta_{\rm Ax}$ is the in-plane Alfv\'en speed~$\beta_{\rm Ax} = \big[ \sigma/[1 + \sigma (1 + b_g^2)] \big]^{1/2}$. 
    As~$\beta_i \to 0$ (i.e., for particles with initial velocity nearly perpendicular to the outflow), the resulting energy gain becomes
    \begin{equation} \label{eq:W_Fermi}
    W_{\rm Fermi}/m_ec^2 = \lim_{\beta_i \to 0} \big( 1 - (\Delta \beta)^2 \big)^{-1/2} - 1 = \frac{1 + \beta_{\rm Ax}^2}{1 - \beta_{\rm Ax}^2} - 1 = \frac{2\sigma}{1 + \sigma b_g^2}.
    \end{equation}
    Hence a strong guide field (i.e., $b_g \gg \sigma^{-1/2}$) significantly damps Fermi kicks, resulting in an energization of~$W_{\rm Fermi} \simeq 2 b_g^{-2} \ll 2\sigma$. Conversely, a weak guide field ($b_g \ll \sigma^{-1/2}$) allows efficient energization with~$W_{\rm Fermi} \simeq 2\sigma$.

    \item [(c)] \textbf{Pickup acceleration}: An incoming upstream particle crosses the reconnection separatrix suddenly and thus (i) the particle's magnetic moment in the outflow frame~$\mu' \equiv \gamma' \lvert \textbf{p}'_\perp \rvert^2 / 2 \lvert \textbf{B}' \rvert$ is no longer constant, resulting in a greater perpendicular momentum~$\lvert \textbf{p}'_\perp \rvert$ and (ii) the particle becomes immersed in a reconnection outflow with bulk ($E\times B$-drift) velocity of $v_{\rm out} \sim v_{\rm Ax}$. The ideal-MHD (i.e., motional) electric field~$\textbf{E}_m \equiv - \textbf{u} \times \textbf{B}$ subsequently accelerates the particle until its perpendicular guiding-center velocity matches the $E \times B$-drift velocity of the outflow (Figure~\ref{fig:inj_mech_cartoon}c) \citep{Drake2009,French_2023,Chernoglazov2023}. 
    
    Thus the total work done by this pickup mechanism on a particle of initial energy~$\gamma_0$ is 
    \begin{equation} \label{eq:W_pickup}
        W_{\rm pickup}/m_ec^2 \equiv \gamma_{\rm Ax} - \gamma_0 = \sqrt{1 + \frac{\sigma}{1 + \sigma b_g^2}} - \gamma_0,
    \end{equation}
    where~$\gamma_{\rm Ax} = \left( 1 - \beta_{\rm Ax}^2\right)^{-1/2}$. 
\end{enumerate}
Several strategies have been employed in the literature to disentangle the energetic contribution of each mechanism to particle energization.
One common approach utilizes the motional (i.e., ideal) electric field~$\textbf{E}_m$ and the non-ideal electric field~$\textbf{E}_n \equiv \textbf{E} - \textbf{E}_m$ by approximating $\textbf{E}_n$ as the electric field component parallel to the local magnetic field, $\textbf{E}_n \simeq \textbf{E}_\parallel \equiv (\textbf{E} \cdot \textbf{B})\textbf{B}/\lvert \textbf{B} \rvert^2$, and $\textbf{E}_m$ as the perpendicular component, $\textbf{E}_m \simeq \textbf{E}_\perp \equiv \textbf{E} - \textbf{E}_\parallel$; one then compares the works~$W_\parallel$ and~$W_\perp$ done by these electric-field components over a statistically large ensemble of tracer particles \citep{Guo2019,Comisso2019,Kilian2020,French_2023}. 
Since the value of the normalized guide-field strength adopted in our present study, $b_g = 0.3$, exceeds the dimensionless reconnection rate~$\eta_{\rm rec} \simeq 0.1$ \citep{Guo2015,Liu2015,Liu2017,Liu2020,Werner2018,Goodbred_2022,HaoZhang_2023}, which quantifies the typical strength of the downstream reconnected in-plane magnetic field relative to the upstream reconnecting magnetic field, we shall proceed with this approximation.

Thus, in this work we compute the \textit{relative contributions of each mechanism to the total injected particle population} (i.e., ``injection shares") by subjecting each tracer particle in the ensemble to the following scheme \citep{French_2023}. Upon reaching~$\gamma_{\rm inj}$ (i.e.,  when the particle is ``injected"), the particle is endowed with its ``injection time" $t_{\rm inj} \equiv W^{-1}(\gamma_{\rm inj})$, where~$W(t)$ is the total time-evolved work done to the particle and~$W^{-1}(\gamma)$ is its inverse function. Then the particle is assigned to a single mechanism according to whichever of the following conditions is true at~$t = t_{\rm inj}$:
\begin{equation} \label{eqn:mechanisms}
\begin{gathered}
(W_\parallel > W_\perp) \ \& \ (\lvert \textbf{p}_\parallel \rvert > \lvert \textbf{p}_\perp' \rvert) \implies \ \text{$E_{\rm rec}$ acceleration} \\
(W_\perp > W_\parallel) \ \& \ (\lvert \textbf{p}_\parallel \rvert  > \lvert \textbf{p}_\perp' \rvert) \implies \ \text{Fermi kick(s)} \\
(W_\perp > W_\parallel) \ \& \ (\lvert \textbf{p}_\perp' \rvert > \lvert \textbf{p}_\parallel \rvert) \implies \ \text{Pickup process},
\end{gathered}
\end{equation}
with the remaining possibility categorized as ``other." Each~$\textbf{p}_\parallel$ is left unprimed because boosting to the~$E\times B$-drift frame makes no change to momenta parallel to the local magnetic field.

\section{Simulation setup} \label{sec:setup}

To study particle injection and acceleration by relativistic magnetic reconnection, we perform an array of fully kinetic simulations of a relativistic collisionless pair plasma using the \code{Zeltron} code, which solves the relativistic Vlasov-Maxwell equations \citep{Cerutti2013} in 2D and 3D rectangular domains. All of our simulations are initialized with a force-free current sheet (CS) with the associated initial magnetic field
\begin{equation} \label{eq:mag_field}
   \textbf{B} \equiv B_0 \tanh{(y/\lambda)} \ \hat{\textbf{x}} + B_0 \sqrt{\sech^2{(y/\lambda)} + b_g^2} \  \hat{\textbf{z}}, 
\end{equation}
where~$\lambda = \sqrt{3} \, \sigma \rho_0$ is the half-thickness of the initial CS and~$\rho_0 \equiv c/\omega_{\rm ce} = m_e c^2/e B_0$ is the nominal relativistic gyroradius.

The pair mass ratio is~$m_i/m_e = 1$. The initial plasma density~$n_0 = n_e + n_i$ is uniform and represented by~$8$ ($32$) positron-electron pairs per computational grid cell in 3D (2D), as justified in Appendix~\ref{sec:convergence_studies}. Currents are normalized by~$J_0 \equiv en_0c/2$. The initial plasma is thermal with a uniform temperature~$\theta_0 \equiv T_0/m_ec^2 = 0.3$. The guide-field strength is set to~$b_g = 0.3$.

To examine the effects of magnetization and dimensionality, our simulations vary these quantities independently. In particular, we scan the cold upstream magnetization parameter over eight values: $\sigma \in \{8, 12, 16, 24, 32, 48, 64, 96\}$, where the latter three values are exclusive to 2D due to limited computational resources. Since~$\theta_0$ is sub-relativistic, these~$\sigma$ values are close to their corresponding ``hot" upstream magnetizations~$\sigma_h \simeq \sigma/\Gamma_{\rm th}$. 
To examine the effect of dimensionality we compare 2D and 3D simulations that are otherwise identical. 

In this study, we characterize system sizes by the dimensionless measure~$\ell \equiv L/\sigma \rho_0$, where~$L$ is the system size. The spatial domains of our simulations are rectilinear boxes with~$x \in [0, \ell_x]$ and~$y \in [-\ell_y/2, \ell_y/2]$, and in 3D also with~$z \in [0, \ell_z]$. The aspect ratio is fixed to~$\ell_x = \ell_y = 2\ell_z$ ($\ell_x = \ell_y$) in 3D (2D). To resolve kinetic scales, the spatial resolution is set to~$\Delta x = d_e/2$, as justified in Appendix \ref{sec:convergence_studies} (where~$d_e \equiv c/\omega_{\rm pe} = \sqrt{2\sigma} \rho_0$ is the relativistic collisionless electron skin depth). The temporal resolution is set to~$\Delta t = \omega_{\rm pe}^{-1}/6$. In the~$x$ and~$z$-directions, periodic boundary conditions are set for fields and particles, whereas in the~$y$-direction conducting boundaries are set for fields and reflecting boundaries are set for particles. Instead of adding a small perturbation to trigger magnetic reconnection, we instead wait for the current sheet to start reconnecting spontaneously.\footnote{Previous work has found that introducing a perturbation with~$b_g = 0$ significantly influences 3D results; in particular, making them more similar to 2D results which are insensitive to a perturbation \citep{Werner2021}. However, when performing a similar test using a guide magnetic field of strength~$b_g = 0.3$, we find that the 3D results (energy evolution, injection shares, particle spectra, and reconnection rates) are insensitive to a perturbation, possibly because of the stabilizing effect of the guide field that already makes 3D results more similar to 2D.}  

The simulation domain size is fixed to~$\ell_x = 256 = 128 \sqrt{2\sigma} \, d_e/\sigma \rho_0 \simeq 148 \, \lambda/\sigma \rho_0$, to yield results that are converged in large-system-size limit (informed by our previous work; cf. \citet{French_2023}). Correspondingly, we use~$N_x \in$ $\{ 1024,$ $1280,$ $1440,$ $1792,$ $2048,$ $2560,$ $2880,$ $3584 \}$ computational grid cells in the~$x$-direction across the~$\sigma$ scan, where accordingly the latter three values are exclusive to 2D. 
The running time varies for each simulation, but is set to allow enough time for the plasma to (a) reconnect and (b) evolve for~$\geq 3\, L_x/c = 768 \, \sigma \omega_{\rm ce}^{-1} \simeq 543 \, \sqrt{\sigma} \omega_{\rm pe}^{-1}$ after reconnection onsets. This duration is sufficient for evolving quantities to saturate or otherwise reach steady-state evolution.

In this paper, we wish to investigate particle spectra in the ``downstream," i.e., the region between the two separatrices. To isolate the downstream region, we apply the particle-mixing approach \citep{Daughton2014} as follows. Each computational grid cell is assigned a mixing fraction~$\mathcal{F}_e \equiv (n_e^{\rm bot} - n_e^{\rm top}) / (n_e^{\rm bot} + n_e^{\rm top})$, where~$n_e^{\rm bot}$ and~$n_e^{\rm top}$ are the number densities of electrons that start at~$y < 0$ and~$y > 0$, respectively. Then, we consider cells which satisfy~$\lvert \mathcal{F}_e \rvert \leq 70\% (96\%)$ to be sufficiently well-mixed to be regarded as ``downstream" in 3D (2D) and the remaining cells are considered ``upstream" in 3D (2D).\footnote{By contrast, in 3D particles may escape the downstream and, from the perspective of the mixing diagnostic, proceed to ``contaminate" the upstream, leading to the small portions of the upstream being mis-identified as downstream. To combat this issue a lower mixing threshold may be used, which in turn however may mis-identify part of the downstream as upstream. Considering these two issues, a recent study has found the optimal mixing ratio threshold in 3D to be 70\% \citep{Hao_2021}.} 

When evaluating the contributions of each injection mechanism (see Section~\ref{ss:result_inj_shares}), we randomly select~$10^6$ particles at the beginning of each simulation and track relevant physical quantities associated with them at each time step, including velocities and electric and magnetic fields. 
With this information, we study the acceleration mechanisms of particles statistically \citep{Guo2016,Guo2019,Li2019a,Li2019b,Kilian2020,French_2023}. We exclude the initial current-carrying (i.e., drifting) particles from our tracer particle analysis. 
\section{Results} \label{sec:results}

\begin{figure}
    \centering
    \includegraphics[width=\textwidth]{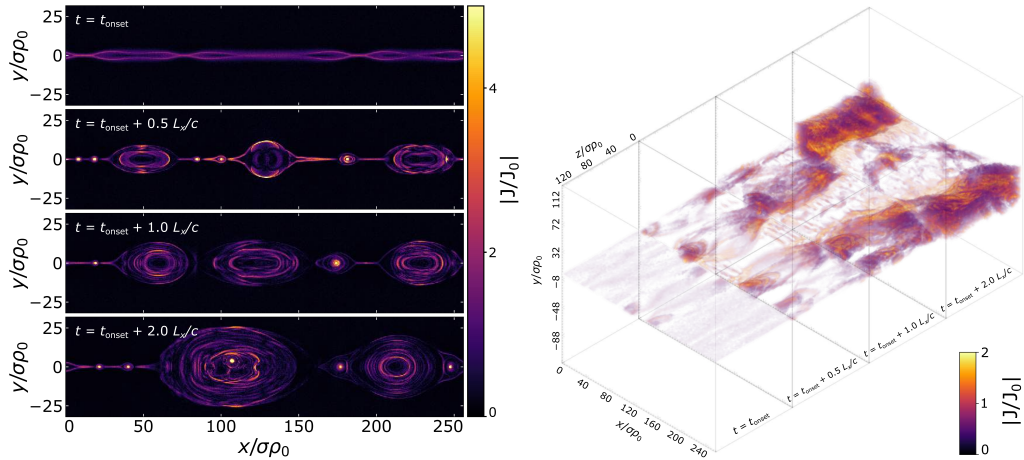}
    \caption{Absolute current density~$\lvert J/J_0 \rvert$ at different times after the time of reconnection onset~$t_{\rm onset}$. Panels on the left display a 2D simulation ($\sigma = 8$) and the panels on the right display an otherwise identical 3D simulation.}
    \label{fig:abs_J}
\end{figure}

First, to illustrate the reconnection process, we show several snapshots of absolute current density~$\lvert J/J_0 \rvert$ (where~$J_0 \equiv en_0c/2$ and~$J$ is the total current density) of two simulations with~$\sigma = 8$ (Figure~\ref{fig:abs_J}). The left-hand panels display a 2D simulation and the right-hand panels display an otherwise identical 3D simulation. To make the downstream visible in 3D, we apply a linear ramp of increasing opacity for~$J \geq J_0/2$ (opacity = 50\% at~$J = J_0/2$ and opacity = 100\% at~$J = 2 \, J_0$) and set opacity = 0 for~$J < J_0/2$.

The earliest time displayed is the reconnection onset time~$t_{\rm onset}$, defined as when the magnetic energy drops to~$99.99\%$ of its initial value. This threshold is sufficient for~$t_{\rm onset}$ to demarcate the very first X-point collapse. Henceforth we will shift to this time for time-dependent results. 

Accordingly, initial X-point collapse in 2D is shown to coincide with~$t_{\rm onset}$ in the top-left panel of Figure~\ref{fig:abs_J}. Soon thereafter, plasmoids are advected downstream ($t_{\rm onset} + 0.5 \,L_x/c$), merge with other plasmoids ($t_{\rm onset} + 1.0 \,L_x/c$), and retain their structural integrity throughout ($t_{\rm onset} + 2.0 \,L_x/c$). By contrast, in 3D the flux-rope kink instability dismantles plasmoids, allowing particles to escape from them \citep{Dahlin2014,Qile_2021,Werner2021}.

\subsection{Reconnection rate} \label{ss:result_rec_rate}

We define the dimensionless time-dependent reconnection rate~$\eta_{\rm rec}(t)$ as the rate at which the unreconnected (i.e., upstream) magnetic flux $\psi_{\rm up}$ decays with time, normalized by $v_{\rm A0} B_0$:
\begin{equation}  \label{eqn:etarec_timedep}
    \eta_{\rm rec}(t) \equiv -\frac{1}{v_{\rm A0} B_0} \expval{\frac{\partial \psi_{\rm up}}{\partial t}}, \  \psi_{\rm up}(t) \equiv \frac{1}{L_x L_z} \int_{\rm up} B_x(t) \,dV, \ v_{\rm A0} \equiv c \sqrt{\frac{\sigma}{1 + \sigma}}, 
\end{equation}
where~$\expval{}$ is the time averaging performed every~$(1/32) \, L_x/c$ and the upstream region is defined by cells with a mixing fraction~$\lvert \mathcal{F}_e \rvert > 70\%$ ($96\%$) in 3D (2D). The peak reconnection rate is defined simply as~$\text{max}[\eta_{\rm rec}(t)]$.

\begin{figure}
    \centering
    \includegraphics[width=\textwidth]{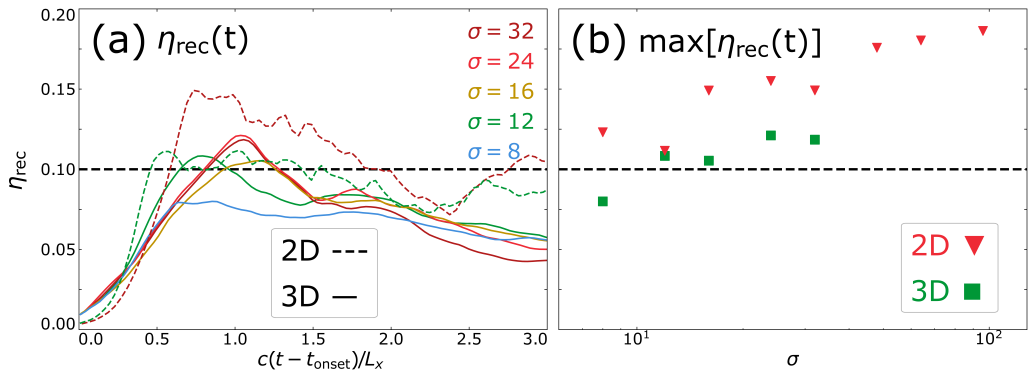}
    \caption{Reconnection rates for various~$\sigma$. (a): Time-dependent reconnection rates of 3D (solid) and a few 2D (dashed) simulations. (b): Peak reconnection rates, with green squares representing 3D and red triangles representing~2D.
    }
    \label{fig:etarec_vs_sigma_and_time}
\end{figure}

In accordance with Eq.~\eqref{eqn:etarec_timedep}, we compute the time-dependent and peak reconnection rates for several values of~$\sigma$, as shown in Figure~\ref{fig:etarec_vs_sigma_and_time}. We find that the reconnection rate in 3D is consistently somewhat lower than each 2D counterpart for every tested value of~$\sigma$, and adheres quite closely to the canonical value of~$0.1$. As for~$\sigma$-dependence, the peak reconnection rate~$\text{max}[\eta_{\rm rec}(t)]$ increases gradually with increasing~$\sigma$, from~$\simeq 0.1$ to~$0.2$ in 2D as~$\sigma$ varies from~$8$ to~$96$ and from~$\simeq 0.08$ to~$0.12$ in 3D as~$\sigma$ varies from~$8$ to~$32$.

\subsection{Injection energies} \label{ss:result_inj_energies}

\begin{figure}
    \centering
    \includegraphics[width=\textwidth]{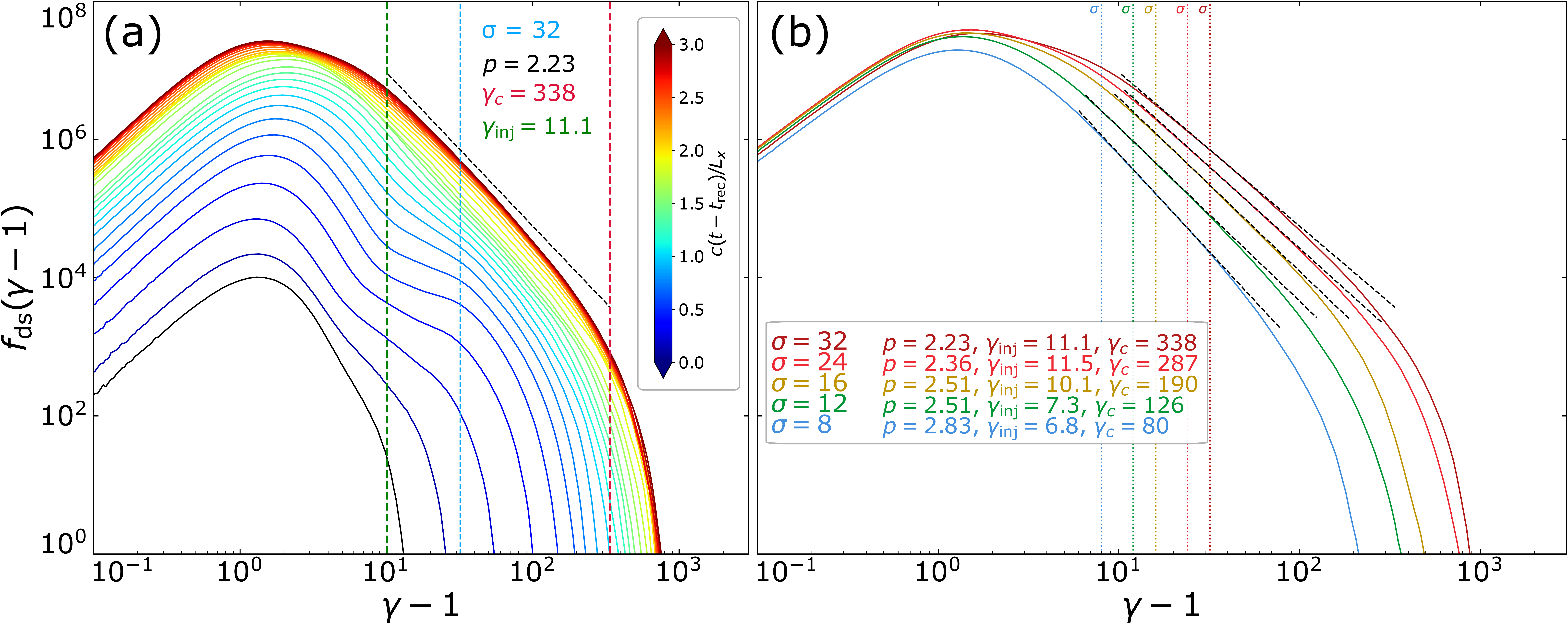}
    \caption{Downstream particle spectra from 3D simulations. 
    Panel~(a): Evolving downstream particle spectrum from a~$\sigma=32$ 3D simulation fitted at the final time step. The vertical dashed green line indicates the measured injection energy~$\gamma_{\rm inj}$, the vertical dashed red line the measured cutoff energy~$\gamma_c$, and the dashed black line is~$\gamma^{-p}$ with measured power-law index~$p$. 
    Solid color lines show particle spectra taken every~$(1/8) \, L_x/c$, from~$t = t_{\rm onset}$ to~$t = t_{\rm onset} + 3\, L_x/c$. 
    Panel~(b): Downstream particle spectra of 3D simulations at times~$t = t_{\rm onset} + 3\, L_x/c$ for various initial upstream magnetizations~$\sigma$. Dashed black lines show~$\gamma^{-p}$ for~$\gamma \in [\gamma_{\rm inj}, \gamma_c]$ using the measured values of~$p, \gamma_{\rm inj}, \gamma_c$ and dotted vertical lines are colored and positioned at~$\sigma$ values.}
    \label{fig:spectrum_example}
\end{figure}

To fit particle spectra, we perform a spectral fitting procedure described in Appendix~\ref{sec:Appendix_procedure}. To illustrate the process for acquiring the characteristic spectral parameters, we show a time-evolved downstream particle spectrum~$f_{\rm ds}(\gamma - 1, t)$ in Figure~\ref{fig:spectrum_example}a. The downstream particle population is continuously fed by upstream particles crossing over the separatrix. 

Measuring the characteristic parameters~$p, \gamma_{\rm inj}, \gamma_c$ helps glean their dependencies on system parameters. In Figure~\ref{fig:spectrum_example}b, we show several downstream particle spectra of 3D simulations at the time~$t = t_{\rm onset} + 3\, L_x/c$ for various values of~$\sigma$, as well as power-law fits that extend from the injection energy to the cutoff energy. This figure highlights the precision of measurement that is afforded by implementing the spectral fitting procedure described in Appendix~\ref{sec:Appendix_procedure}.

\begin{figure}
    \centering
    \includegraphics[width=\textwidth]{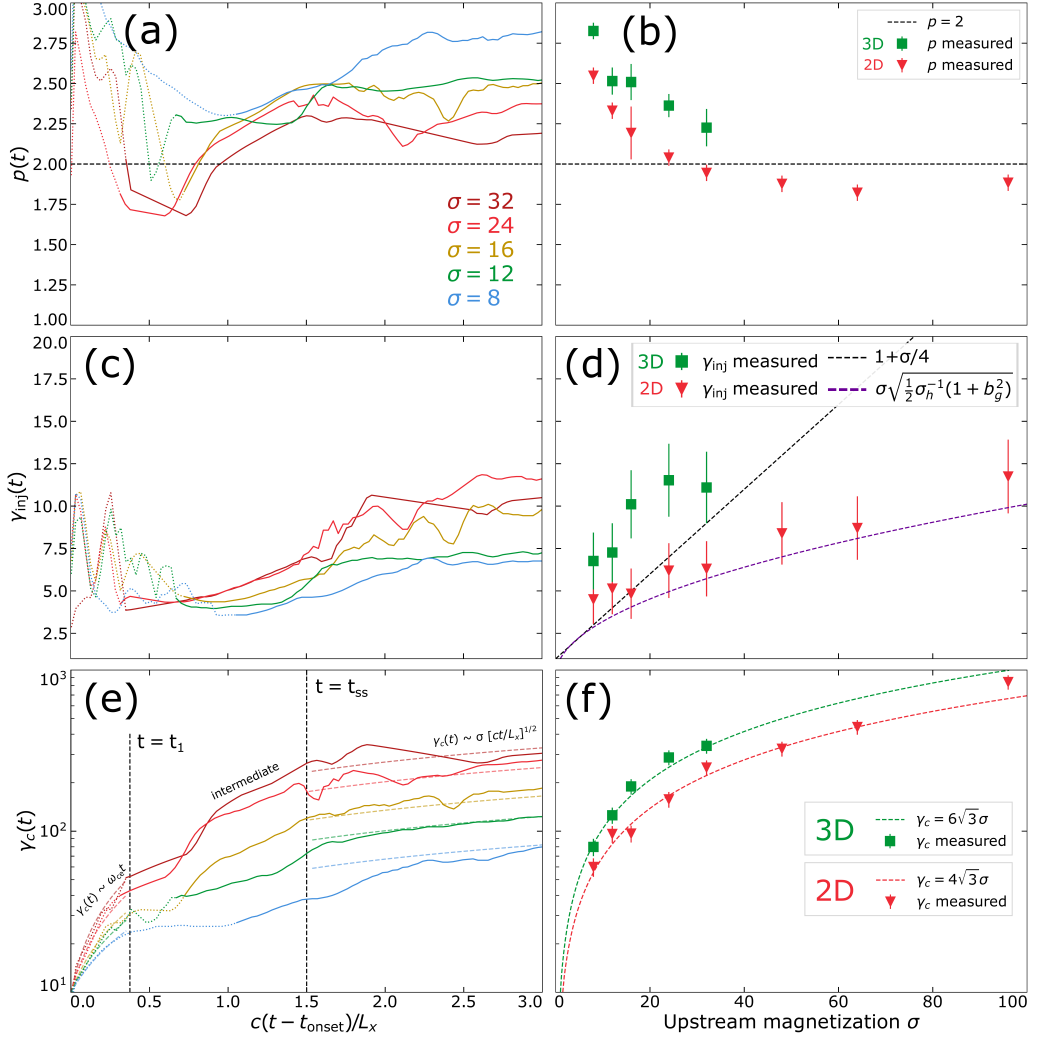} 
    \caption{Spectral parameters measured via fitting procedure (described in Appendix~\ref{sec:Appendix_procedure}) for various~$\sigma$ at each time step for 3D simulations (panels~a, c, e) and at~$t = t_{\rm onset} + 3\, L_x/c$ for all simulations (panels~b, d, f). Dotted colored lines in panels~(a, c, e) indicate time steps where the power-law extent is short, i.e., $\gamma_c/\gamma_{\rm inj} < 10$. 
    In panels~(b, d, f), red triangles are 2D runs and green squares are 3D runs. 
    Panels~(a, b): Power-law indices~$p(t)$ and~$p(t_{\rm onset} + 3\,L_x/c)$. 
    Panels~(c, d): Injection energies~$\gamma_{\rm inj}(t)$ and~$\gamma_{\rm inj}(t_{\rm onset} + 3\,L_x/c)$. The dashed black line in~(d) shows linear scaling, assuming~$\gamma_{\rm inj} = 1 + \sigma/4$ and the dashed purple line shows~$\gamma_{\rm inj} \simeq \sigma \big[ \sigma_h^{-1} \, (1 + b_g^2)/2 \big]^{1/2}$ [i.e., Eq.~\eqref{eqn:ginj_model_soft}], derived in Section~\ref{ss:injection_energy_model}.
    Panels~(e, f): High-energy cutoffs~$\gamma_c(t)$ and~$\gamma_c(t_{\rm onset} + 3\,L_x/c)$. The semi-transparent dashed colored lines in~(e) show the fit from Eq.~\eqref{eqn:gc_fit} and the green (red) dashed line in~(f) shows it evaluated at~$t_{\rm onset} + 3\,L_x/c$, i.e. $\gamma_c(\sigma) = 6\sqrt{3}\sigma$ in 3D ($\gamma_c(\sigma) = 4\sqrt{3}\sigma$ in~2D). 
    }
    \label{fig:params_vs_sigma_and_time}
\end{figure}

In the lefthand panels of Figure~\ref{fig:params_vs_sigma_and_time} we show the evolved characteristic parameters of the power-law particle spectra ($p, \gamma_{\rm inj}, \gamma_c$) for various values of~$\sigma$ represented by curves of different color. Data points are obtained every~$(1/32) \, L_x/c$ using the fitting procedure described in Appendix~\ref{sec:Appendix_procedure} and we smooth the data with a moving time-average of window size~$(3/32) \, L_x/c$, i.e., where each value is replaced by the average of itself, its predecessor, and its successor. In the righthand panels of Figure~\ref{fig:params_vs_sigma_and_time} we plot each measured parameter evaluated at~$t = t_{\rm onset} + 3 \, L_x/c$ against~$\sigma$. 

We find that power-law indices~$p(t)$ (top row of Figure~\ref{fig:params_vs_sigma_and_time}) rapidly harden during the transient phase~($t < t_1$), followed by a longer period whereupon they gradually soften, achieving stability~$\sim 2$ light-crossing times after~$t_{\rm onset}$ (panel~a). Power-law spectra harden with increasing~$\sigma$ and are harder in 2D than in 3D by~$\sim 0.2$-$0.3$ for any given value of~$\sigma$ (panel~b).

The injection energy~$\gamma_{\rm inj}(t)$ (middle row of Figure~\ref{fig:params_vs_sigma_and_time}) quickly adopts an initial (i.e., at~$t = t_{\rm onset}$) value of~$\sim 6 \pm 2$ without a clear dependence on~$\sigma$. While erratic, the injection energy remains roughly within this range during the transient reconnection phase [$0 \leq t - t_{\rm onset} < (3/8) \, L_x/c \simeq 100 \, \sigma \omega_{\rm ce}^{-1}$]. 
Once steady state is reached (e.g., after~$1.5 \, L_x/c \simeq 400 \, \sigma \omega_{\rm ce}^{-1}$ in 3D), the injection energy stabilizes roughly on the same timescale with which~$p(t)$ stabilizes (panel~c). At~$t_{\rm onset} + 3 \, L_x/c$, the measured injection energies in 2D adhere reasonably closely to our model for the injection energy in the thermally-dominated case~$\gamma_{\rm inj} \simeq \sigma \big[ \sigma_h^{-1} (1 + b_g^2)/2 \big ]^{1/2}$ [cf., Eq.~\eqref{eqn:ginj_model_soft}] (dashed purple line) and is greatly exceeded by the often-assumed linear relation of~$\gamma_{\rm inj} = \sigma/4$ (indicated by the dashed black line).

In 3D, injection energies are consistently greater by a factor of~$\sim 1.5$ than in~2D, possibly owed to greater current sheet thicknesses at disruption in 3D vs 2D (cf., Section~\ref{ss:injection_energy_model}). 
However, the large errors and the limited covered range of~$\sigma$ makes a definite scaling trend difficult to decipher. We compare these results with previous work in Section~\ref{ss:ginj_w_sigma}.

Although not the primary concern of this study, we measure the high-energy cutoff $\gamma_c$ to grow rapidly during the transient reconnection phase $0 \leq t - t_{\rm onset} \lesssim t_1 = (3/8) \, L_x/c \simeq 100 \, \sigma \omega_{\rm ce}^{-1}$, in both 2D and~3D. By assuming~$\gamma_c(t)$ to scale with~$W_{\rm direct}(t) \sim \eta_{\rm rec} \omega_{\rm ce} t$, we apply the fit~$\gamma_c(t) = a \, \omega_{\rm ce} (t - t_{\rm onset})$, where~$a$ is a fitting parameter. Performing this fit in 3D yields the coefficients~$a = 0.020, 0.017, 0.015, 0.014, 0.014$ for~$\sigma = 8, 12, 16, 24, 32$. The sensitivity of~$a$ to~$\sigma$ declines as the~$\sigma \gg 1$ regime is entered. These fits are plotted on Figure (\ref{fig:params_vs_sigma_and_time}e). 

The steady-state phase ($t - t_{\rm onset} \geq t_{\rm ss}$, where $t_{\rm ss}$ is the steady-state time) is characterized by the time interval when the simulation domain is populated with multiple plasmoids and the cutoff grows steadily. In 3D, the transition from the transient to the steady-state phase takes~$\sim 1 \, L_x/c$ to complete for~$\sigma \gg 1$, with longer transition times at moderate values of~$\sigma$ (cf. $\sigma = 8, 12$). Thus~$t_{\rm ss} = t_{\rm onset} + 1.5 L_x/c \simeq t_{\rm onset} + 400 \, \sigma \omega_{\rm ce}^{-1}$. 
During the steady-state phase ($t > t_{\rm ss}$), $\gamma_c(t)$ is fitted reasonably well by~$\gamma_c(t) = 6 \sigma \big[ c(t - t_{\rm onset})/L_x \big]^{1/2}$ for every value of~$\sigma$ in 3D (panel~e). 
Thus the complete cutoff energy fit in 3D is (where~$a \simeq 0.014$ in the~$\sigma \gg 1$ limit):
\begin{equation} \label{eqn:gc_fit}
    \gamma_c(t) = \begin{cases}
    a \, \omega_{\rm ce}(t - t_{\rm onset}) & 0 \leq t - t_{\rm onset} < t_1 \\ 
    \text{intermediate} & t_1 \leq t - t_{\rm onset} < t_{\rm ss} \\
    6 \sigma \big[ c(t - t_{\rm onset})/L_x \big]^{1/2} & t - t_{\rm onset} \geq t_{\rm ss}
\end{cases} \end{equation}

By contrast, in 2D the transition from transient to steady-state is much more sudden (not shown). The complete cutoff energy fit in 2D is (where~$a \simeq 0.014$ in the~$\sigma \gg 1$ limit):
\begin{equation} \label{eqn:gc_fit_2D}
   \gamma_c(t) = \begin{cases}
   a \, \omega_{\rm ce}(t - t_{\rm onset}) & 0 \leq t - t_{\rm onset} < t_1 \simeq t_{\rm ss} \\ 
   4 \sigma \big[ c(t - t_{\rm onset})/L_x \big]^{1/2} & t - t_{\rm onset} \geq t_{\rm ss}
\end{cases} \end{equation}

We also explicitly measure the high-energy cutoff at~$t = t_{\rm onset} + 3 \, L_x/c$ and find that it indeed adheres well to a linear scaling with~$\sigma$ (panel~f). We discuss the coefficient~$a$ and comparisons to previous work in Section~\ref{ss:gc(t)}.

\subsection{Injection efficiencies} \label{ss:result_inj_efficiencies}

\begin{figure}
    \centering
    \includegraphics[width=\textwidth]{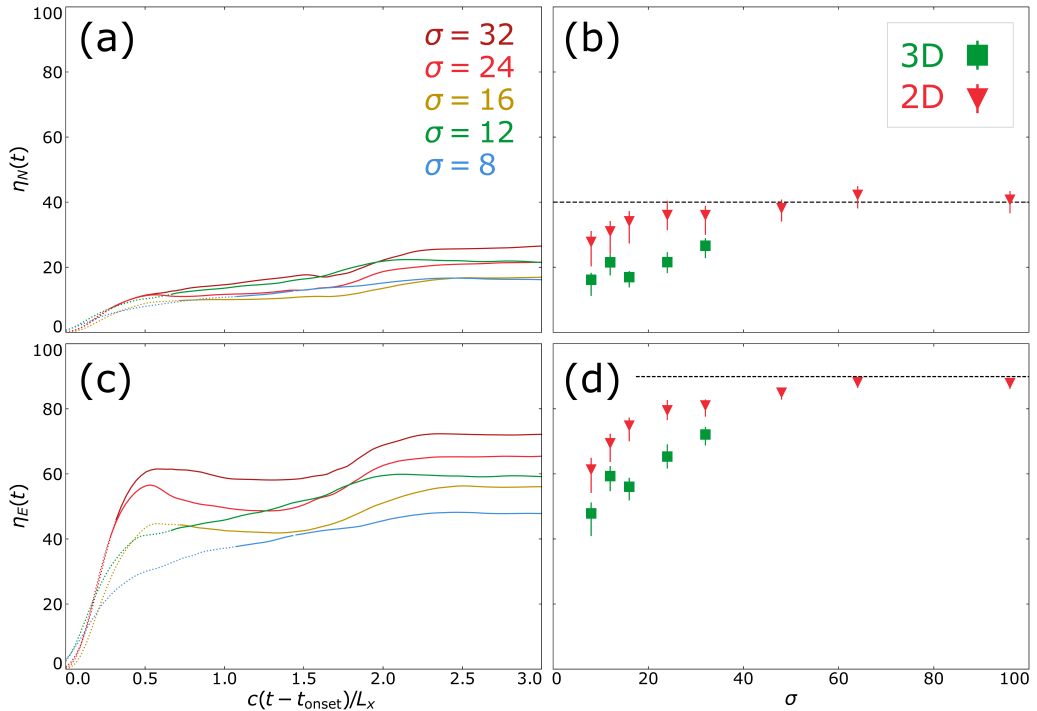} 
    \caption{Efficiencies of particle injection and energy computed for 3D simulations at various~$\sigma$ at each time step (panels a, c) and at final times (panels b, d), comparing 2D (red triangle) and 3D (green square) runs. Dotted segments in panels~(a, c) indicate time steps for which the power-law extent was short (i.e., $t$ for which~$\gamma_c(t)/\gamma_{\rm inj}(t) < 10$), whereas solid lines indicate times where~$\gamma_c(t)/\gamma_{\rm inj}(t) > 10$. Panels~(a, b): Injection efficiencies, where the dashed black horizontal line indicates~$\eta_{N} = 40\%$ on panel~(b). Panels~(c, d): Energy efficiencies, where the dashed black horizontal line indicates~$\eta_{E} = 90\%$ on panel~(d). Time-evolved errors are not shown but are comparable to the final-time errors.}
    \label{fig:efficiencies_f_and_evolution}
\end{figure}

To contextualize the injection shares in the broader downstream population, we compute the injection (or number) efficiency, defined by 
\begin{equation} \label{eq:eta_N}
    \eta_{N}(t) \equiv N_{\rm inj}(t)/N_{\rm ds}(t), \ N_{\rm inj}(t) \equiv \int_{\gamma_{\rm inj}(t)}^\infty f_{\rm ds}(\gamma, t) \,d\gamma, \ N_{\rm ds}(t) \equiv \int_1^\infty f_{\rm ds}(\gamma, t) \,d\gamma,
\end{equation}
and the energy efficiency, defined by
\begin{equation} \label{eq:eta_E}
    \eta_{E}(t) \equiv E_{\rm inj}(t)/E_{\rm ds}(t), \ E_{\rm inj}(t) \equiv \int_{\gamma_{\rm inj}(t)}^\infty \gamma f_{\rm ds}(\gamma, t) \,d\gamma, \ E_{\rm ds}(t) \equiv \int_1^\infty \gamma f_{\rm ds}(\gamma, t) \,d\gamma.
\end{equation}
Since these quantities depend on~$\gamma_{\rm inj}(t)$ measured from time-evolving particle spectra, these efficiencies are well-defined for~$t > t_{\rm onset}$, i.e. when power-law spectra may be deciphered by our  spectrum fitting procedure. 

We use Eqs.~\eqref{eq:eta_N},~\eqref{eq:eta_E} to directly compute~$\eta_N(t)$ and~$\eta_E$(t) for various~$\sigma$, as shown in Figure~\ref{fig:efficiencies_f_and_evolution}. 

We generally find that both~$\eta_N(t)$ (panel~a) and~$\eta_E(t)$ (panel~c) display a period of rapid initial growth for about~$0.5 \, L_x/c$ after~$t_{\rm onset}$, slowing down significantly at intermediate times, eventually saturating at a finite~$\sigma$-dependent values~$\eta_{N,\rm sat}(\sigma)$ and~$\eta_{E,\rm sat}(\sigma)$ at late times, $t-t_{\rm onset} \gtrsim 2 \, L_x/c$, in agreement with the slower late-time growth of~$\gamma_c(t)$ and stabilization of~$p(t)$. 
In 2D, the time-saturated injection efficiency $\eta_{N,\rm sat}$ improves from~$\sim 30\%$ to~$\sim 40\%$ as $\sigma$ is increased from~$8$ to~$96$, at which point convergence with~$\sigma$ is achieved (panel~b). 
The 3D simulations systematically show somewhat lower values of~$\eta_{N,\rm sat}$ (20-25\%) than their 2D counterparts for any given value of~$\sigma$, likely owed to softer power-law indices; cf. Fig.~\ref{fig:params_vs_sigma_and_time}. 
However, because of the limited $\sigma$-range covered by our 3D simulation campaign, it is difficult to extract any clear and reliable systematic trends for the dependence of~$\eta_{N,\rm sat}$ on~$\sigma$ in~3D.

The final saturated energy efficiency~$\eta_{E,\rm sat}(\sigma)$ in 2D simulations also grows monotonically starting from~$\sim 60\%$ $\sigma=8$ and saturates at the ~$\sim 90\%$-level for $\sigma \gtrsim 50$ (panel~d). In~3D, $\eta_{E,\rm sat}(\sigma)$ is again lower than in~2D, but, unlike $\eta_{N,\rm sat}(\sigma)$, exhibits clear monotonic growth with~$\sigma$, exceeding 70\% at the highest probed value $\sigma=32$. Extending our 3D study to even higher values of~$\sigma$ is clearly needed in order to determine how and what level~$\eta_{E,\rm sat}(\sigma)$ saturates in the ultra-relativistic limit~$\sigma \to \infty$. 

In summary, the time-converged injection and energy efficiencies become high ($\eta_N \simeq 40\%$, $\eta_E \simeq 90\%$) in 2D and are insensitive to~$\sigma$ once it is sufficiently great. While these efficiencies are systematically lower in 3D likely due to softer power-law indices, they are not yet converged at~$\sigma = 32$.

\subsection{Injection shares} \label{ss:result_inj_shares}

\begin{figure}
    \centering
    \includegraphics[width=\textwidth]{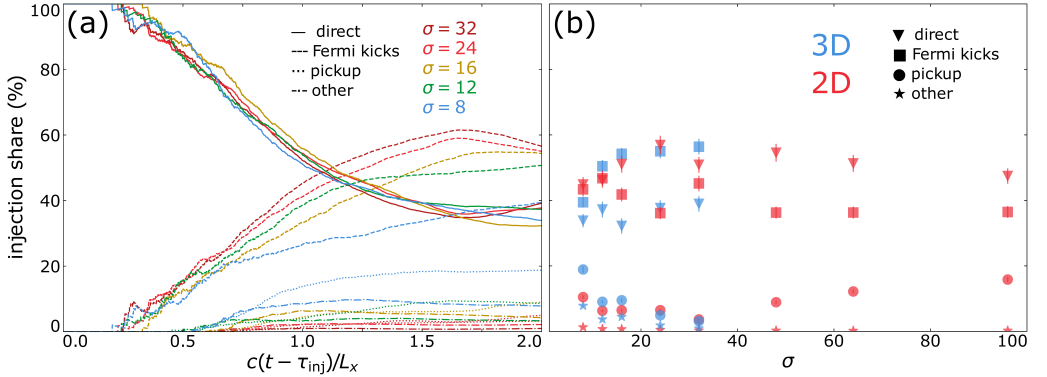} 
    \caption{Contributions of each injection mechanism to the injected particle population. (a): Time-evolved injection shares from 3D runs for several values of~$\sigma$. (b): Injection shares over all runs at~$t = \tau_{\rm inj} + 2\, L_x/c$. 
    All injection shares have an error of~$\sim 3\%$ for each mechanism at every time step, propagated from errors of~$\gamma_{\rm inj}$.} 
    \label{fig:injection_shares}
\end{figure}

To obtain the time-evolved contribution of each injection mechanism (see Section~\ref{ss:injection_mechanisms}) to the total population of injected particles, we apply the formula
\begin{equation} \label{eqn:inj_share_formula}
    \mathcal{S}_i(t) \equiv \frac{\mathcal{N}^{\rm inj}_i(t)}{\sum_j \mathcal{N}^{\rm inj}_j(t)}
\end{equation}
where~$\mathcal{N}^{\rm inj}_i(t)$ is the cumulative number of particles injected by mechanism~$i$ at time~$t$. We term these~$\mathcal{S}_i(t)$'s ``injection shares" and plot them in Figure~\ref{fig:injection_shares} against~$c(t-\tau_{\rm inj})/L_x$, where~$\tau_{\rm inj} \gtrsim t_{\rm onset}$ is the moment when the first injection of a tracer particle occurs. 

In Panel~(a), we show time-evolved injection shares for 3D simulations for the scanned values of~$\sigma$. Here, we find a specific activation sequence of the injection mechanisms that is consistent with our conceptual picture of injection in Figure~\ref{fig:inj_mech_cartoon}. Soon after~$t_{\rm onset}$, the~$W_{\rm direct}$ share is at~$100\%$. At~$t \simeq \tau_{\rm inj} + 0.25 \, L_x/c$, Fermi kicks are activated and gradually (over~$1 \, L_x/c$) become the dominant injection mechanism (except for~$\sigma = 8$, where the pickup process takes a significant~$\sim 20\%$ fraction). Finally, at~$t \simeq \tau_{\rm inj} + 0.5 \, L_x/c$, particles start getting injected by the pickup mechanism, roughly coincident with the time when reconnection outflows are established. The time intervals between the activation of each subsequent mechanism appear independent of~$\sigma$. While not shown in this Figure, we also calculate the fraction of injected particles that are injected at these activation times: at~$c(t - \tau_{\rm inj})/L_x = 0.25, 0.5, 1$, only~$\sim 1/1000$, $\sim 1/100$, and~$\sim 1/10$ of particles that end up injected by~$t = \tau_{\rm inj} + 2 \, L_x/c$ have been injected. This shows that the early dominance of direct acceleration may be misleading, since it represents~$< 10\%$ of the aggregate population of injected particles.

In Panel~(b), we plot the injection shares at~$t = t_{\rm onset} + 2\, L_x/c$ against~$\sigma$ for 2D and 3D simulations. We find that injections by Fermi and direct acceleration remain competitive: Aside from~$\sigma = 8$, we find in 3D that Fermi kicks contribute~$\sim 50\%$ whereas direct acceleration contributes~$\sim 40\%$, and that these shares are flipped in 2D. Meanwhile, when increasing~$\sigma$ from~$8$ to~$32$, we find that pickup shares are suppressed significantly, from~$20\%$ ($10\%$) $\to 5\%$ in 3D (2D); however, they curiously rise back up in 2D when varying~$\sigma$ from~$48$ to~$96$. 
A negligible~($\sim 1 \%$) fraction of particles satisfy~$(W_\parallel > W_\perp) \, \& \, (\lvert \textbf{p}_\perp' \rvert > \lvert \textbf{p}_\parallel \rvert)$ upon reaching the injection energy~$\gamma_{\rm inj}$ (categorized as ``other" in Section~\ref{ss:injection_mechanisms}), suggesting that any mechanisms which impart this combination onto particles can be ignored.

\subsection{NTPA correlations of each mechanism} \label{ss:result_NTPA_correlations}

Given a particle of energy~$\gamma \geq \gamma_{\rm inj}$, what is the probability that it was injected by the mechanism~$\mathcal{M}_i$? 
In other words, to what extent is each injection mechanism likely, by mere correlation, to produce particles of a given energy~$\gamma$ (e.g., $\gamma \gg \gamma_{\rm inj}$)? We attempt to answer this question by introducing the ``NTPA correlation" $\mathcal{C}_i$ of each injection mechanism~$\mathcal{M}_i$, defined by
\begin{equation} \label{eq:ntpa_corr}
    \mathcal{C}_i (\gamma) \equiv \frac{\mathcal{N}_i(\gamma, \tau)}{\sum_i \mathcal{N}_i (\gamma, \tau)},
\end{equation}
where~$\mathcal{N}_i(\gamma, \tau)$ is the number of particles injected by mechanism~$\mathcal{M}_i$ with energy~$\gamma \geq \gamma_{\rm inj}$ at a certain time~$\tau$ that will be taken to be~$\tau = \tau_{\rm inj} + 2\, L_x/c$. We term these quantities ``correlations" to emphasize that the $\mathcal{C}_i$'s merely represent the correlation between high-energy particles and their injection mechanism of origin.

\begin{figure}
    \centering
    \includegraphics[width=\textwidth]{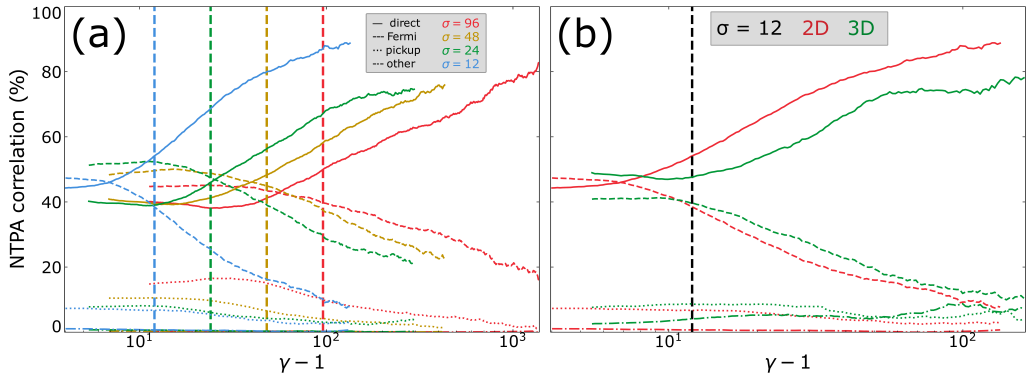}
    \caption{NTPA correlation of each injection mechanism evaluated at~$\tau = \tau_{\rm inj} + 2\, L_x/c$ plotted against~$\gamma - 1$ for~$\gamma \in [\gamma_{\rm inj}, \gamma_c]$, with~$\sigma$ values indicated by the vertical dashed lines. (a): NTPA correlations of four 2D simulations with~$\sigma = 12, 24, 48, 96$ indicated by blue, green, gold, and red lines respectively. (b): NTPA correlations of 2D (red) and 3D (green)~$\sigma = 12$ simulations.
    }
    \label{fig:NTPA_correlations}
\end{figure}

To calculate~$\mathcal{C}_i(\gamma)$, we define~$\lfloor 5\, \gamma_c/\gamma_{\rm inj} \rfloor$ (i.e., rounded to the nearest integer below~$5\, \gamma_c/\gamma_{\rm inj}$) energy bins for~$\gamma \in [\gamma_{\rm inj}, \gamma_c]$ spaced uniformly in log space. Then, we populate these bins with the final energies particles attain, sorted by which mechanism injected them; this yields~$\mathcal{N}_i(\gamma, \tau)$. Finally, we obtain~$\mathcal{C}_i(\gamma)$ by normalizing each histogram by~$\sum_i \mathcal{N}_i (\gamma, \tau)$, thereby ensuring that~$\sum_i \mathcal{C}_i(\gamma) = 100\%$ at each energy bin. In Figure~\ref{fig:NTPA_correlations} we plot~$\mathcal{C}_i(\gamma)$ for each injection mechanism for~$\gamma \in [\gamma_{\rm inj}, \gamma_c]$.  

\begin{figure}
    \centering
    \includegraphics[width = 8.5cm]{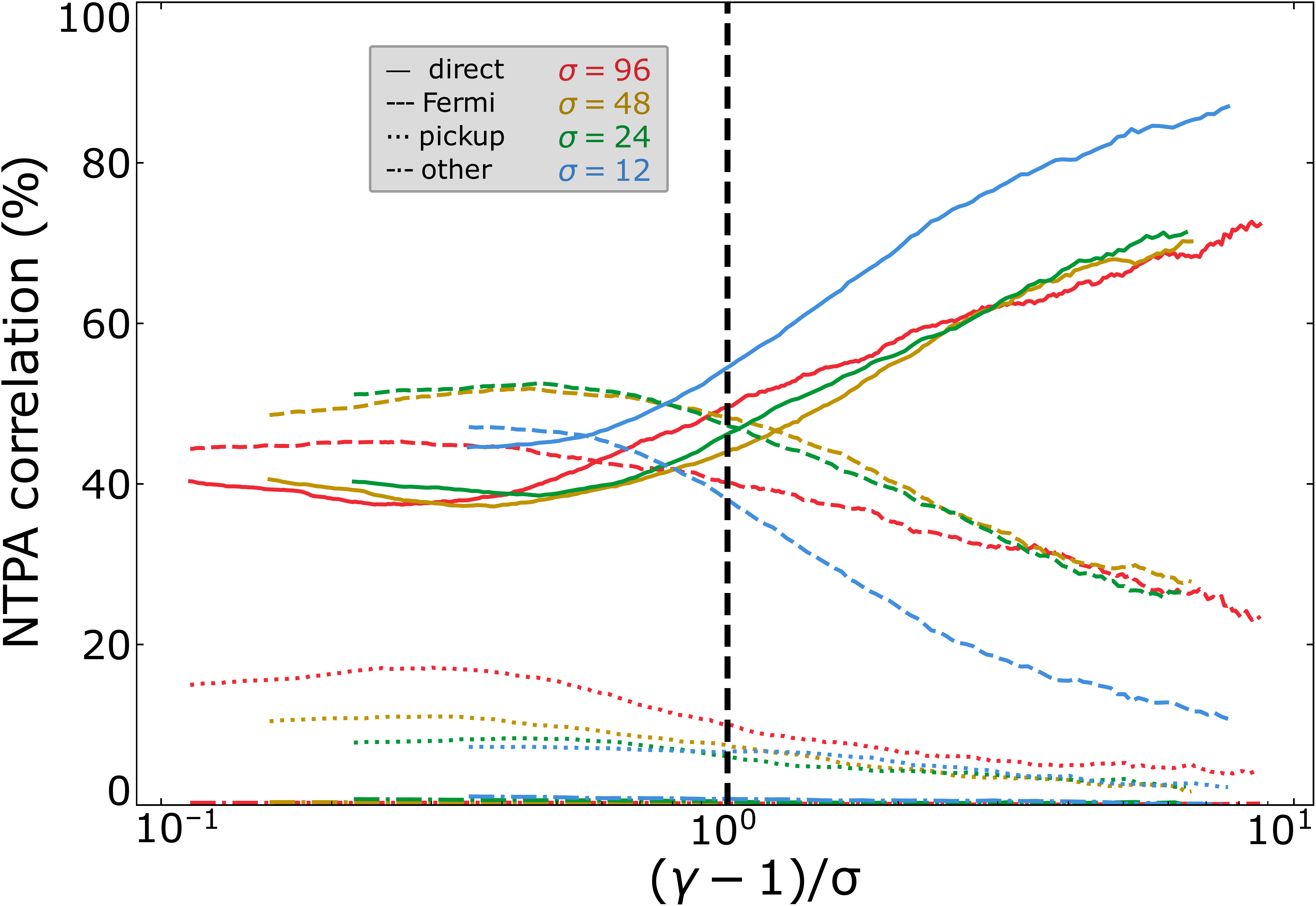}
    \caption{NTPA correlation of each injection mechanism evaluated at~$\tau = \tau_{\rm inj} + 2\, L_x/c$ plotted against~$(\gamma - 1)/\sigma$ for~$\gamma \in [\gamma_{\rm inj}, \gamma_c]$ and~$\sigma \in [12, 24, 48, 96]$. The black dashed vertical line indicates~$\gamma - 1 = \sigma$.}
    \label{fig:NTPA_correlations_vs_gs}
\end{figure}

In Figure~\ref{fig:NTPA_correlations}a, we compare the effects of magnetization across four 2D simulations with $\sigma = 12, 24, 48, 96$. We find that the more energetic a given particle is at the final time, the greater the likelihood that it was injected by~$W_{\rm direct}$, with~$\sim 80\%$ of particles around~$\gamma = \gamma_c$ having been injected by this mechanism. Furthermore, it appears that~$\mathcal{C}_i(\gamma, \sigma) = \mathcal{C}_i(\gamma/\sigma)$. To see this more clearly, we have plotted this separately in Figure~\ref{fig:NTPA_correlations_vs_gs}. 

In Figure~\ref{fig:NTPA_correlations}b, we compare 2D and~3D. Adding a third dimension appears to slightly dampen the direct acceleration correlations, with~$\sim 75\%$ of particles at~$\gamma = \gamma_c$ being injected by direct acceleration rather than~$\sim 90\%$ at~$\sigma = 12$.

In summary, despite direct acceleration injecting only~$\simeq 40$-$50\%$ of particles, there is a~$\sim 80\%$ NTPA correlation between the particles with~$\gamma \sim \gamma_c$ and injection by direct acceleration. These results may explain some of the disparities in the literature on particle injection in the weak guide field ($b_g \lesssim 0.3$) regime, wherein \citet{Totorica_2023} and \citet{Gupta_2025} found direct acceleration to dominate NTPA correlations and \citet{French_2023} found Fermi kicks to dominate injection shares. We elaborate further in Section~\ref{ss:4_NTPA_correlations}.

\section{Discussion} \label{sec:discussion}

\subsection{The injection criterion} \label{ss:ginj_w_sigma}

Historically, the meaning of the term \textit{injection} in the published literature on relativistic magnetic reconnection has varied. The works by \citet{Ball2019}, \citet{Kilian2020}, and \citet{French_2023} consider injection to be the energization of particles up until the low-energy bound of the power-law spectrum. Meanwhile, \citet{Sironi2022}, \citet{Totorica_2023}, and \citet{Gupta_2025} describe injection as the ``early stages" of particle acceleration. Nevertheless, each concept has been made rigorous in a similar fashion by:
\medskip
\noindent
\begin{enumerate}[label=(\roman*), leftmargin=30pt, itemindent=0pt, labelsep=10pt]
    \item defining ``injection" as the acceleration that occurs up to a threshold energy~$\gamma_{\rm inj}$ \citep{Ball2019,Kilian2020,Sironi2022,French_2023,Totorica_2023,Gupta_2025}, and

    \item modeling this threshold as~$\gamma_{\rm inj} = 1 + \sigma$ \citep{Ball2019,Kilian2020,Sironi2022} or~$\gamma_{\rm inj} = 1 + \sigma/4$ \citep{Sironi2022,Totorica_2023,Gupta_2025}.
\end{enumerate} 

\medskip
While we [in the present work and in \citet{French_2023}] retained convention (a), i.e., that there exists a certain characteristic energy~$\gamma_{\rm inj}$, surpassing which can serve as the criterion for injection, we discarded assumption~(b), opting instead to measure~$\gamma_{\rm inj}$ from particle spectra directly. 
Our procedure for measuring~$\gamma_{\rm inj}$ (cf., Appendix~\ref{sec:Appendix_procedure} and \citet{French_2023}) is guided by the first conception mentioned above, i.e., injection as the energization of particles up until the low-energy bound of the nonthermal power-law spectrum. 

Our measurements in the present work (where~$b_g = 0.3$ and~$\theta_0 = 0.3$) have revealed that the~$\gamma_{\rm inj} = \sigma$ model vastly overestimates the low-energy bound of the power-law spectrum in both 2D and 3D (Figure~\ref{fig:params_vs_sigma_and_time}d). In 2D, the~$\gamma_{\rm inj} = 1 + \sigma/4$ model still overestimates measurements for $\sigma \gtrsim 30$; however, in 3D we lack the data at such high values of~$\sigma$ to make the same conclusion. 

The 2D measurements are, however, consistent with our model for the injection criterion, which defines a particle as \textit{injected} if its gyroradius~$r_g$ exceeds the elementary current sheet half-thickness~$\delta/2 \simeq d_{\rm e, rel}/2 = c/(2\omega_{\rm pe, rel})$ (cf. Section~\ref{ss:injection_energy_model}). In the thermally-dominated regime, this yields an injection energy~$\gamma_{\rm inj} \simeq \sigma \left[ \sigma_h^{-1} \, (1 + b_g^2)/2 \right]^{1/2}$.
In particular, this model is consistent with our measurements of the upstream magnetization dependence of~$\gamma_{\rm inj}$ in 2D (Figure~\ref{fig:params_vs_sigma_and_time}d). In 3D, however, the model systematically underestimates the measurements, which we speculate is owed to current sheets disrupting at greater thicknesses in 3D. 

A potentially important implication of this model is that if no particles can achieve~$r_g \geq \delta/2$, then no particles are injected and thus a power-law spectrum of nonthermal particles does not form. Therefore, if one can determine \textit{a priori} whether any particles can be injected, one could determine precisely, from system parameters alone, the conditions under which a power-law spectrum can form. A future study could test this systematically by (a) varying~$\sigma_h$ from~$\sigma_h \lesssim 1$ to~$\sigma_h \gg k^{-1}$, (b) estimating the injection energy and the work done by each mechanism over this parameter range, and (c) seeing whether the formation of a power-law spectrum coincides with the ability for at least one mechanism to inject particles. Moreover, a future study could examine the effects of using an electron-ion plasma, where the two-tiered structure of current sheets may have a significant effect on the injection energy and relevant injection mechanisms for each species.

While the present study focuses on a single guide-field strength~$b_g = 0.3$, the injection criterion developed in Section~\ref{ss:injection_energy_model} provides insight into the expected guide-field dependence of~$\gamma_{\rm inj}$. The explicit factor of~$\left( 1 + b_g^2 \right)^{1/2}$ in Eq.~\eqref{eqn:ginj_model_proper} implies that stronger guide fields increase the injection energy, as the enhanced total magnetic field strength reduces the particle gyroradius and thus requires greater particle energy to satisfy the criterion~$r_g \geq \delta/2$. For weak guide fields ($b_g \ll 1$), $\gamma_{\rm inj}$ is insensitive to~$b_g$, whereas for strong guide fields ($b_g \gtrsim 1$), $\gamma_{\rm inj}$ grows approximately linearly with~$b_g$. This scaling is roughly consistent with measurements in \citet{French_2023}, which focused on 2D reconnection and found that~$\gamma_{\rm inj}/\sigma$ increases from~$\sim 0.15$ at~$b_g = 0.1$ to~$\sim 0.30$ at~$b_g = 1.0$. A systematic study that compares measurements of~$\gamma_{\rm inj}$ over varying both~$b_g$ and~$\sigma$ in 3D relativistic reconnection remains a direction for future work.

\subsection{The injection shares and NTPA correlations} \label{ss:4_NTPA_correlations}

In addition to the injection criterion having differing conceptions in the literature (Section~\ref{ss:ginj_w_sigma}), there have also been differing conceptions about how to evaluate the relative importance of various injection mechanisms. In this work and \citet{French_2023}, we consider \textit{injection shares}---i.e., the fractional contribution of each mechanism to the total injected particle population---to reflect that mechanism's importance. In contrast, works by \citet{Totorica_2023} and \citet{Gupta_2025} essentially\footnote{The assessment of mechanism contributions done in \citet{Totorica_2023} and \citet{Gupta_2025} differs from the \textit{NTPA correlations} defined in Section~\ref{ss:result_NTPA_correlations} only in that, instead of quantifying the contributions by the discrete metric of the number of particles they each inject, in those works they are quantified by the overall amount of work they do on particles with~$\gamma < \gamma_{\rm inj}$. This is possible because these studies consider mechanisms characterized by time-integrated quantities [i.e., work done by ideal vs non-ideal electric fields in \citet{Totorica_2023} and work done by $E<B$ vs $E>B$ electric fields in \citet{Gupta_2025}]. For a statistically large sample of tracer particles, the shares defined by the number of particles injected by a mechanism should approach those defined in terms of the overall work done on particles with energies~$\gamma < \gamma_{\rm inj}$ by that mechanism. } 
consider NTPA correlations, evaluated at the highest permissible energies, to reflect mechanism importance. 

Despite these studies evaluating different mechanisms, we may nevertheless make a comparison with the NTPA correlations evaluated in this study. Indeed, we also find that the higher the final particle energy, the greater the likelihood (approaching~$\sim 80\%$) that the particle  was injected by direct acceleration, which is often associated with non-ideal (including $E>B$) electric fields when~$b_g \sim 0$ (Figures~\ref{fig:NTPA_correlations}, \ref{fig:NTPA_correlations_vs_gs}). However, while NTPA correlations at high energies [e.g., $\mathcal{C}_i(\gamma = \gamma_c)$] can certainly be quite useful in their own right, they have little bearing on the dominant injection mechanism. 
This is because particles with a head start in entering the power law---such as those injected by direct acceleration, which dominates at early times as suggested by Figure~\ref{fig:injection_shares}---will participate in the continual Fermi mechanism that operates in the power law for a longer duration, allowing them to achieve the highest energies. This significantly biases~$\mathcal{C}_i(\gamma = \gamma_c)$ as a metric for evaluating particle injection, since the particles injected at times closer to~$t_{\rm onset}$ are over-represented. 

Conversely, the injection shares can be read off Figure 2b of~\citet{Totorica_2023} by evaluating each line at~$\epsilon_{\rm final} = \epsilon^*$, which yields~$\textbf{E}_n$ contributing~$\sim 30\%$ to the injected particle population for the threshold $\epsilon^* = \sigma/4$ for reconnection with~$\sigma = 50, \, b_g = 0.125$; this is somewhat consistent with our previous work which found~$W_\parallel$ to contribute~$\sim 20\%$ for the threshold $\gamma_{\rm inj} \simeq 1 + \sigma/6$ when~$\sigma = 50, \, b_g = 0.1$~\citep{French_2023}. Likewise, the injection shares can be read off Figure~3 of~\citet{Gupta_2025} by evaluating each line at~$\epsilon_T/\sigma = \epsilon^*$, which shows that~$E>B$ electric fields inject~$\sim 5\%$ of particles, in rough agreement with~\citet{Guo2022_comment}. 

The guide-field strength is also expected to affect the relative contributions of the injection mechanisms. As shown in Eqs.~\eqref{eq:W_Fermi} and~\eqref{eq:W_pickup}, both Fermi kicks and pickup acceleration are suppressed by factors containing~$(1 + \sigma b_g^2)$ in the denominator. Consequently, for weak guide fields ($b_g \ll \sigma^{-1/2}$), these mechanisms associated with the motional (perpendicular) electric field can efficiently inject particles, whereas for strong guide fields ($b_g \gtrsim 1$), their energization is significantly damped. In the strong guide-field regime, direct acceleration by the parallel electric field---which is not suppressed by the guide field---is expected to dominate the injection process. This trend is consistent with \citet{French_2023}, which found that~$W_\parallel$ contributions increase from~$\sim 20\%$ to~$\sim 70\%$ as~$b_g$ increases from~$0.1$ to~$1.0$.

\subsection{The high-energy cutoff \texorpdfstring{$\gamma_c$}{}} \label{ss:gc(t)}
Power-law spectra are often modeled with an exponential cutoff, i.e. $f(\gamma) \sim \gamma^{-p}\, e^{-\gamma/\gamma_c}$, where~$\gamma_c$ is the \textit{high-energy cutoff}. We will discuss here two questions about this quantity: (a) How does~$\gamma_c$ evolve with time? (b) How does~$\gamma_c$ vary with upstream magnetization~$\sigma$? 

Let us begin with the first question. The time evolution of~$\gamma_c$ can be roughly decomposed into three phases: (i) the short, transient phase of rapid rise which immediately follows reconnection onset ($t_{\rm onset} < t < t_1$), (ii) an intermediate phase, characterized by a mixed reconnection layer that is populated with both current sheets and developing plasmoids ($t_1 < t < t_{\rm ss}$), and (iii) the slow, quasi-steady phase, characterized by fully developed multi-plasmoid dynamics in the reconnection layer ($t > t_{\rm ss}$). We show how~$\gamma_c$ evolves during these phases in 3D for several values of~$\sigma$ in Figure~\ref{fig:params_vs_sigma_and_time}e (Section~\ref{ss:result_inj_energies}). 

To our knowledge, no dedicated study has yet numerically explored the evolution~$\gamma_c(t)$ systematically during phase (i). Our high-cadence depositions of particle spectra have enabled us to explore this for the first time in detail. We find that this phase lasts approximately~$(3/8) \, L_x/c \simeq 100 \, \sigma \omega_{\rm ce}^{-1}$; our measurements of~$\gamma_c(t)$ adhere closely to the simple linear form~$\gamma_c(t) = a\, \omega_{\rm ce}(t - t_{\rm onset})$ in both 2D and~3D, where~$a = 0.017 \pm 0.003$ is a fitting parameter. This scaling of~$\gamma_c(t)$ agrees with the time-dependent work done by the reconnection electric field, $W_{\rm direct}(t) = \eta_{\rm rec} \omega_{\rm ce} t \, m_e c^2$, except somewhat smaller in magnitude [$\gamma_c(t) \simeq W_{\rm direct}/6 m_e c^2$]. This agreement conveys a narrative consistent with the injection shares, which suggest that direct acceleration in small, elementary diffusion regions around X-points is the exclusive injector of particles during the first~$0.25 \, L_x/c$ following reconnection onset (Figure~\ref{fig:injection_shares}). The discrepancy of~$\sim 1/6$ may be attributed to the fact that particles with finite drift in the $x$-$y$ plane will leave elementary diffusion regions around X-points on microscopic timescales. 

The existence of an intermediate phase in 3D where~$\gamma_c(t) \sim t^{r}$ for~$0.5 < r < 1.0$ may be attributed to particles that have escaped plasmoids (e.g., disrupted by the flux-rope-kink instability) and enter Speiser-like orbits which energize them as~$\gamma \sim t$ \citep{Uzdensky2011b,Cerutti2012,Cerutti2014,Dahlin2017,Hao_2021}. As plasmoids develop in the reconnection layer, the relative population of ``trapped" ($\gamma \sim \sqrt{t}$) to ``free" ($\gamma \sim t$) particles increases, eventually leading to the steady-state phase wherein~$\gamma_c \sim \sqrt{t}$. 

During the subsequent quasi-steady phase (i.e., $t > t_{\rm ss}$), previous 2D studies by \citet{Petropoulou2018,Hakobyan2021} discovered that the high-energy cutoff grows as~$\gamma_c \sim \sqrt{t}$ due to the adiabatic compression of plasmoids. This scaling of~$\gamma_c \sim \sqrt{t}$ has also been found in 3D in the transrelativistic~$\sigma_h = 1$ regime by \citet{Werner2021}. Our results reaffirm this scaling in both 2D and 3D, where the measured~$\gamma_c(t > t_{\rm ss})$ adheres closely to the model~$\gamma_c(t > t_{\rm ss}) \simeq 4 \sigma \big[ c(t - t_{\rm onset})/L_x \big]^{1/2}$ in 2D (not shown) and~$\gamma_c(t > t_{\rm ss}) \simeq 6 \sigma \big[ c(t - t_{\rm onset})/L_x \big]^{1/2}$ in 3D (Figure~\ref{fig:params_vs_sigma_and_time}e). Our work in 3D does not however access the high-$\sigma$, large-$L_x$ regime wherein secondary power laws of $\gamma^{-2}$ at higher energies have been found \citep{HaoZhang_2023}. 

The second question of how~$\gamma_c$ varies with~$\sigma$ was originally explored by \citet{Werner2016}, who found $\gamma_c \simeq 4\sigma$ in~2D. We also find~$\gamma_c \sim \sigma$ in both 2D and 3D but with a time-dependent coefficient in accordance with \citet{Petropoulou2018} (Figure~\ref{fig:params_vs_sigma_and_time}f). 

Combining our understanding of these two questions allows us to construct a more complete description of the high-energy cutoff evolution in 3D relativistic magnetic reconnection [Eq.~\eqref{eqn:gc_fit}].

\section{Conclusions} \label{sec:conclusions}

The primary goal of this work was to understand energetic particle injection in relativistic magnetic reconnection --- a poorly understood yet essential aspect of reconnection-driven nonthermal particle acceleration. Therefore, we sought to elucidate (i) what is and what sets the injection criterion and (ii) what physical mechanisms promote the particles to satisfy the criterion. 

More concretely, we performed an array of particle-in-cell simulations of a reconnecting relativistic collisionless pair plasma immersed in a weak guide-field of~$b_g = 0.3$. From these simulations, we (i) measured the injection energy~$\gamma_{\rm inj}$ (Section~\ref{ss:result_inj_energies}) and (ii) evaluated the time-dependent contributions of three distinct acceleration mechanisms to the total injected particle population (Section~\ref{ss:result_inj_shares}), and investigated how these quantities depend on the upstream magnetization~$\sigma \in \{ 8, 12, 16, 24, 32, 48, 64, 96\}$ and dimensionality (2D vs~3D). We have also devised a physical model for~$\gamma_{\rm inj}$ (Section~\ref{ss:injection_energy_model}) which is consistent with 2D results. 

This work also had the secondary goals of (i) evaluating the injection and energy efficiencies~$\eta_N$ and ~$\eta_E$ (Section~\ref{ss:result_inj_efficiencies}) and (ii) uncovering the NTPA correlations of each mechanism, i.e., the correlations between injected particle energy and injection mechanism of origin (Section~\ref{ss:result_NTPA_correlations}). We summarize our findings in the rest of this section.

\subsection{Characteristic spectral parameters and efficiencies}

Using a novel procedure for fitting power-law spectra, we have been able to acquire the three key spectral parameters---power-law index~$p$, injection energy~$\gamma_{\rm inj}$, and cutoff energy~$\gamma_c$ [from~$f \sim \gamma^{-p} e^{-\gamma/\gamma_c}, \gamma \in [\gamma_{\rm inj}, \infty)$]---with unprecedented precision (see Appendix~\ref{sec:Appendix_procedure} for details). This has enabled us to assert the following conclusions: 

\medskip
\noindent
\begin{enumerate}[label=(\roman*), leftmargin=30pt, itemindent=0pt, labelsep=10pt]

    \item \textbf{Injection energy~$\boldsymbol \gamma_{\rm inj}$:} 

    The injection energy~$\gamma_{\rm inj}$ (i.e., the energy at which the power-law component of the downstream particle spectrum begins) grows sub-linearly with increasing magnetization~$\sigma$. In particular, $\gamma_{\rm inj} \simeq 5$ to~$12$ when varying~$\sigma = 8$ to~$96$ in 2D and~$\gamma_{\rm inj} \simeq 7$ to~$12$ when varying~$\sigma = 8$ to~$32$ in~3D; $\sim 50\%$ greater than 2D measurements for each value of~$\sigma$. 

    The 2D measurements of~$\gamma_{\rm inj}$ adhere well to the model we developed for the injection criterion, specifically in the thermally-dominated regime (i.e., the intermediate~$\sigma_h$ regime whereupon the plasma temperature inside elementary current sheets is governed by the ambient upstream plasma), wherein the injection energy is~$\gamma_{\rm inj} \simeq \sigma\big[ \sigma_h^{-1} (1 + b_g^2)/2 \big]^{1/2}$(Section~\ref{ss:injection_energy_model}, Figure~\ref{fig:params_vs_sigma_and_time}d). In contrast, in~3D the model consistently underestimates the measurements by a factor of~$\sim 1.5$. 

    \item \textbf{Power-law index~$p$:}
    After reconnection onset, power-law spectra rapidly harden to~$p \lesssim 2$ and then gradually soften, stabilizing at~$t \sim t_{\rm onset}+ 400 \, \sigma \omega_{\rm ce}^{-1}$ (Figure~\ref{fig:params_vs_sigma_and_time}a). 
    At final times (i.e., $t = t_{\rm onset} + 3\,L_x/c$), the power-law index~$p$ declines with increasing~$\sigma$, varying from~$p \simeq 2.5$ to~$1.9$ when varying~$\sigma = 8$ to~$96$ in 2D and~$p \simeq 2.8$ to~$2.2$ when varying~$\sigma = 8$ to~$32$ in~3D. Power-law spectra are consistently steeper by~$0.2$-$0.3$ in 3D compared to 2D for the same value of~$\sigma$ (Figure~\ref{fig:params_vs_sigma_and_time}b).

    \item \textbf{Cutoff energy~$\boldsymbol{\gamma_c}$:}
    During the first~$\sim 65 \, \sigma \omega_{\rm ce}^{-1}$ following reconnection onset, virtually the only injection mechanism that is active is direct acceleration by~$E_{\rm rec}$ in non-ideal reconnection diffusion regions (cf., Figure~\ref{fig:injection_shares}a).  
    Hence we sought to test whether during its initial rise~$\gamma_c(t)$ would evolve similarly to~$W_{\rm direct}$, i.e. $\dot{\gamma}_c \sim \dot{W}_{\rm direct} = \eta_{\rm rec} \omega_{\rm ce} \simeq 0.1 \, \omega_{\rm ce}$. Indeed, we found that~$\gamma_c(t) = a\, \omega_{\rm ce} (t - t_{\rm onset}) \simeq W_{\rm direct}(t - t_{\rm onset})/6$ closely matches the~$\gamma_c$ measurements during the early transient period~$t - t_{\rm onset} < (3/8) \, L_x/c \simeq 100 \, \sigma \omega_{\rm ce}^{-1}$ [(Figure~\ref{fig:params_vs_sigma_and_time}e)]. 
    After this transient period, we find that~$\gamma_c(t)$ transitions to a slower-growth second stage, adhering closely to~$\gamma_c(t) \simeq 4\sigma \big[ c(t - t_{\rm onset})/L_x \big]^{1/2}$ in 2D and $\gamma_c(t) \simeq 6\sigma \big[ c(t - t_{\rm onset})/L_x \big]^{1/2}$ in~3D, in agreement with previous work (cf. Section~\ref{ss:gc(t)}), see  Figure~\ref{fig:params_vs_sigma_and_time}e.

    \item \textbf{Injection efficiency~$\boldsymbol{\eta_N}$ and energy efficiency~$\boldsymbol{\eta_E}$:} 
       
    The ``efficiencies" of injection~$\eta_N \equiv N_{\rm inj}/N_{\rm ds}$ [Eq.~\eqref{eq:eta_N}; i.e., the fraction of the downstream particles that are injected] and energy~$\eta_E \equiv E_{\rm inj}/E_{\rm ds}$ [Eq.~\eqref{eq:eta_E}; i.e., the fraction of the downstream particle kinetic energy carried by injected particles] grows rapidly during the first~$\sim 0.5 \, L_x/c$ that follows reconnection onset [Figure~\ref{fig:efficiencies_f_and_evolution} panels~(a, c)]. Subsequently, these efficiencies continue to grow throughout the simulation, but slower, and after $t - t_{\rm onset} \gtrsim 2 \, L_x/c$ saturate at finite, $\sigma$-dependent values~$\eta_{N,\rm sat}(\sigma)$ and~$\eta_{E,\rm sat}(\sigma)$. 
    
    The late-time saturated injection efficiency~$\eta_{N, \rm sat}(\sigma)$ increases with~$\sigma$ in both 2D and~3D. In particular, in 2D it grows from about 30\% to~40\% as $\sigma = 8$ to~$96$, while in 3D it increases from about 15\% to~25\% as $\sigma = 8$ to~$32$  [Figure~\ref{fig:efficiencies_f_and_evolution} panels~(b, d)]. 
    Likewise, the time-saturated energy efficiency~$\eta_{E,\rm sat}(\sigma)$ also grows with~$\sigma$, increasing from 60\% to~90\% as $\sigma = 8$ to~$96$ in 2D and from 50\% to~70\% as $\sigma = 8$ to~$32$ in~3D.
    
    In 2D, the $\sigma$-convergence of the efficiencies around~$\eta_{N,\rm sat}(\sigma) \simeq 40\%$ and~$\eta_{E,\rm sat}(\sigma) \simeq 90\%$ is realized for~$\sigma \gtrsim 50$, whereas in 3D such convergence is not yet established, likely due to the limited~$\sigma$-range covered by our 3D simulations. These~$\sigma$-converged 2D measurements are in agreement with our previous study \citep{French_2023}. The consistently lower values of~$\eta_{N, \rm sat}(\sigma)$, $\eta_{E, \rm sat}(\sigma)$ in 3D are likely owed to the steeper power-law spectra found in~3D (Figure~\ref{fig:params_vs_sigma_and_time}b). 
    
\end{enumerate}

\subsection{Injection mechanism shares and NTPA correlations}

\begin{enumerate}[label=(\roman*), leftmargin=30pt, itemindent=0pt, labelsep=10pt]
    \item \textbf{Injection shares:}   
    We find a specific activation sequence of the injection mechanisms that agrees with our geometrical picture of injection (Figure~\ref{fig:inj_mech_cartoon}). Denoting the time of the first tracer particle injection as~$t = \tau_{\rm inj} \gtrsim t_{\rm onset}$, the first mechanism to activate is~$W_{\rm direct}$ at~$t = \tau_{\rm inj}$, followed by~$W_{\rm Fermi}$ at~$t \simeq \tau_{\rm inj} + 0.25 \, L_x/c$, followed by~$W_{\rm pickup}$ at~$t \simeq \tau_{\rm inj} + 0.5 \, L_x/c$ (Figure~\ref{fig:injection_shares}a). 
    Importantly, of all the particles that are injected by the end of each simulation, only~$\sim 10\%$ are injected during the initial~$W_{\rm direct}$-dominated phase (i.e., $t \in [\tau_{\rm inj}, \tau_{\rm inj} + L_x/c]$). This implies that while~$W_{\rm direct}$ is initially important, it can nevertheless become sub-dominant when evaluating the aggregate injected particle population over a longer period. 

    The asymptotic, late-time injection shares from Fermi kicks and direct acceleration trend toward $\sim 40\%$ and~$\sim 50\%$ respectively in 3D (2D) with increasing~$\sigma$ (Figure~\ref{fig:injection_shares}b). 
    Pickup shares are suppressed from $20\%$ ($10\%$) to~$5\%$ in 3D (2D) when varying~$\sigma = 8$ to~$32$, but rise again in 2D when varying~$\sigma = 48$ to~$96$.

    \item \textbf{NTPA correlations:}
    To discover the probability that a given particle of high energy~$\gamma \geq \gamma_{\rm inj}$ was injected by a certain mechanism, we compute ``NTPA correlations" (Figure~\ref{fig:NTPA_correlations}). We find that particles injected by~$W_{\rm direct}$ comprise a majority of particles with~$\gamma > \sigma$ and comprise~$\sim 80$-$90\%$ of particles with~$\gamma = \gamma_c$. Introducing a third dimension alters this result, with the correlation between direct acceleration and particles at~$\gamma = \gamma_c$ falling from~$\sim 90\%$ to~$\sim 75\%$ at~$\sigma = 12$. 

\end{enumerate}

This work advances our understanding of particle injection in the relativistic regime of magnetic reconnection in several ways. By systematically comparing 2D and 3D simulations across a broad range of upstream magnetizations, we found that 3D effects increase injection energies by~$\sim 50\%$ and that the injection energy~$\gamma_{\rm inj}$ scales sub-linearly with upstream magnetization~$\sigma$ when the hot upstream magnetization~$\sigma_h$ is moderate. 
The theoretical picture of particle injection presented in this work provides a physical basis for understanding how particles enter continual acceleration within the power-law spectrum produced by magnetic reconnection and is consistent with measurements of injection energies in 2D.   
Moreover, by distinguishing between injection shares (quantifying the contribution of each mechanism to the injected particle population) and NTPA correlations (which mechanisms injected the particles that reach the highest energies), this work reconciles apparent contradictions in the literature on particle injection in relativistic magnetic reconnection. 
Finally, the time-resolved analysis of injection mechanism activation sequences and the precise spectral fitting procedure developed here establish a framework for future studies of particle injection across different plasma conditions. 
These results have direct implications for modeling high-energy emissions from astrophysical reconnection sites, where the existence of a thermally-dominated regime implies that power-law spectra produced by magnetic reconnection in the high~$\sigma_h$ limit can potentially be much broader than previously expected. 

\section*{Acknowledgements} \label{sec:acknowledgements}

O.F. acknowledges support from the National Science Foundation Graduate Research Fellowship under Grant No. DGE 2040434. This work was also supported by the National Science Foundation via grant AST 1903335 and by NASA via grants 80NSSC20K0545, 80NSSC22K0828, and 80NSSC24K0941. This research used resources of the Anvil supercomputer, which is supported by National Science Foundation award No. 2005632 and is a resource of the Rosen Center for Advanced Computing at Purdue University. We would like to thank Fan Guo for helpful discussions.

\section*{Declaration of Interests}

The authors report no conflict of interest.

\newpage
\appendix
\section{Convergence Studies} \label{sec:convergence_studies}

In general when conducting PIC simulations, it is crucial to have robust results, and as such one aims to use a domain size as large as possible. However since one is constrained by the computational resources that exist, it is important to consider the quantities which control the computational time. For a given simulation code, the computational cost~$C$ scales with the number of particle pushes, given by~$C \sim N_x N_y N_z N_t n_{\rm ppc}$, where~$N_x N_y N_z$ is the number of cells in the domain, $N_t$ is the number of pushes, and~$n_{\rm ppc}$ is the number of particles per cell. 

Since we are preserving aspect ratios to~$L_x = L_y = 2L_z$ ($L_x = L_y$) in 3D (2D), the cost reduces to~$C_{\rm 2D} \sim N_x^3 n_{\rm ppc}$ in 2D and ~$C_{\rm 3D} \sim N_x^4 n_{\rm ppc}$ in 3D. In particular, for a given domain size~$L_x$ we may cast~$N_x$ in terms of spatial resolution~$L_x/N_x = \delta/\Delta x$, where~$\delta$ is a local length scale and~$\Delta x$ is the grid length. These scalings motivate using values of~$\delta/\Delta x$ and $n_{\rm ppc}$ that are as low as possible while producing results (e.g., particle spectra, injection shares, injection efficiencies, etc.) which are (a) approximately invariant to increases of these inputs and (b) sensitive to reductions of these inputs. The parameters we have decided to use for these convergence studies are summarized in Table~\ref{tab:convergence_study_parameters}.

\begin{table}
\centering
\caption{Quantities for convergence studies}
\label{tab:convergence_study_parameters}
\begin{tabular}{lll}
\toprule
Quantity & Resolution convergence study & PPC convergence study \\
\midrule
Domain size ($L_x$) & $128 \sigma \rho_0 = 1024 \, d_e$ & $128 \sigma \rho_0 = 1024 \, d_e$ \\
Upstream magnetization ($\sigma$) & $32$ & $32$ \\
\# of cells in~$x$ ($N_x$) & $ \{768, 1024, 1536, 2048, 3072, 4096\}$ & $1536$ \\
\# ptl per cell per species ($n_{\rm ppc}$) & $ \{256, 128, 64, 32, 16, 8 \}$ & $ \{64, 32, 16, 8 \}$ \\
Resolution ($d_e/\Delta x$) & $ \{ 1.5, 2, 3, 4, 6, 8 \}$ & $3$ \\
Guide-field strength ($b_g$) & $0.3$ & $0.3$ \\
Mixing fraction ($\mathcal{F}$) & $5\%$ & $5\%$ \\
\# of tracer particles ($N_{\rm tracers}$) & $2 \times 10^5$ & $2 \times 10^5$ \\
Ambient upstream temp. ($\theta_0$) & $ \{ \frac{1}{2}, \frac{1}{8} \}$ & $\frac{1}{2}$ \\
Thickness of initial CS ($\lambda/\sigma \rho_0$) & $ \{ \sqrt{3}, \sqrt{91}/3 \}$ for~$\theta_0 \in \{\frac{1}{2}, \frac{1}{8}\}$ & $\sqrt{91}/3$ \\
Running time ($c\tau/L_x$) & $ \{ 4, 6\}$ for~$\theta_0 \in \{\frac{1}{2}, \frac{1}{8}\}$ & $4$ \\
Drift velocity ($\beta_d$) & $ \{ 0.5, 0.3\}$ for~$\theta_0 \in \{\frac{1}{2}, \frac{1}{8}\}$ & $0.5$  \\ 
Aspect ratio ($N_x/N_y$) & $1$ & $1$ \\
Number of cells in~$z$ ($N_z$) & $1$ & $1$ \\
\bottomrule
\end{tabular}
\end{table}

To save computational resources, we will conduct these convergence studies in 2D and extend the outcomes to 3D as follows. For resolution convergence, we will assume that the minimum spatial resolution necessary to resolve in 3D is identical to 2D. For~$n_{\rm ppc}$ convergence, we will assume that the number of particles per Debye~$n$-sphere remain fixed, so that~$\big( 2 \, n^{(3D)}_{\rm ppc}\big)^3 = \big( 2 \, n^{(2D)}_{\rm ppc} \big)^2$, allowing us to straightforwardly solve for~$n^{(3D)}_{\rm ppc}$.

\subsection{Resolution convergence study} 

The particular length scale~$\delta$ which must be resolved for converged results is not obvious, and therefore we shall experiment with different scales. In order to satisfy the CFL~\citep{CFL_1928} condition, it is necessary to forbid particles moving at light speed from traversing more than one grid length~$\Delta x$ per time step~$\Delta t$, i.e. $d_e/\Delta x > 1$, where~$d_e \equiv c/\omega_{\rm pe}$ is the collisionless relativistic electron skin depth. Therefore the first length scale we will test is~$d_e$, i.e. we seek to determine whether there exists some critical resolution~$\delta_e^{\rm crit}/\Delta x$ above which results vary negligibly and below which results are sensitive. 

Additionally, we will conduct a parallel set of simulations that use a different initial upstream background temperature~$\theta_0 \equiv k_B T_0/m_ec^2$. This is done to uncover which value of~$\lambda_{\rm De} \equiv \sqrt{\theta_0} \, d_e$ which must be resolved. This is also done to determine whether there exists a regime of~$\theta_0 < 1$ for which all particle injection results are approximately invariant. This is of interest not only to reveal increased generality of results, but also potentially to save computational resources, because (a) using~$\theta_0 = 1/2$ yields larger values~$\lambda_{\rm De}/\Delta x$, (b) it means that a larger drift speed~$\beta_d$ can be implemented in the initial current layer without initiating the two-stream instability, thereby causing reconnection to start earlier, and (c) the initial current sheet is thinner, causing it to occupy a smaller fraction of the total particle population for a given length in the~$y$-direction~$L_y$, thereby yielding more precise results for tracer particle analysis. 

In this (2D) convergence study we preserve the number of particles per Debye volume, so that doubling the resolution folds~$n_{\rm ppc}$ by a factor of 4. 

\begin{figure}
    \centering
    \includegraphics[width=\textwidth]{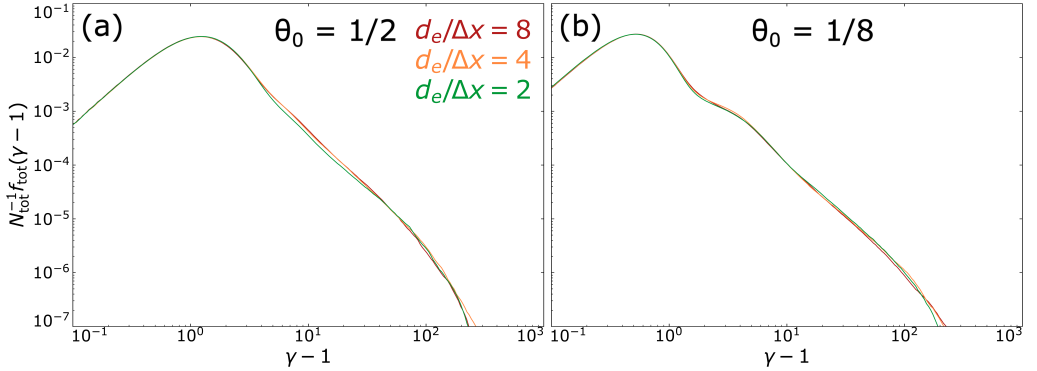}
    \caption{Particle spectra for different resolutions. Panel~(a) uses an ambient upstream temperature of~$\theta_0 = 1/2$ whereas panel~(b) uses~$\theta_0 = 1/8$.}
    \label{fig:res_conv_spectra}
\end{figure}

\begin{figure}
    \centering
    \includegraphics[width=\textwidth]{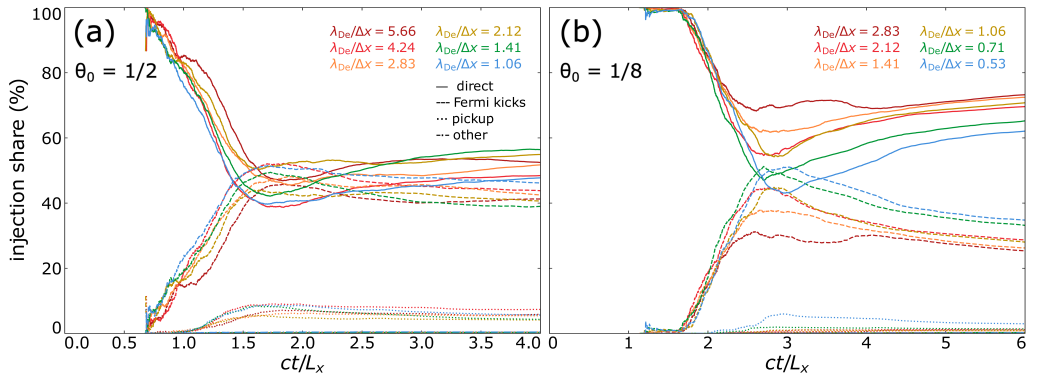}
    \caption{Injection shares for different resolutions. In each panel, the corresponding resolution in skin depths is~$d_e/\Delta x \in \{ 8, 6, 4, 2, 1.5\}$ from dark red to blue lines. Panel~(a) uses an ambient upstream temperature of~$\theta_0 = 1/2$ whereas panel~(b) uses~$\theta_0 = 1/8$.}
    \label{fig:res_conv_inj_shares}
\end{figure}

Regardless of whether the background temperature is~$\theta_0 = 1/8$ or~$1/2$, we find that the particle spectra are invariant of resolution, implying that all resultant spectral quantities ($\gamma_{\rm inj}, \gamma_c, p, \eta_N, \eta_E$) are converged (Figure \ref{fig:res_conv_spectra}). Furthermore, while the injection energy (i.e., the low-energy bound of the power-law spectrum) may be difficult to visually distinguish from the thermal component when~$\theta_0 = 1/2$ (panel~a), it becomes clear when~$\theta_0 = 1/8$ that~$\gamma_{\rm inj} \simeq 5$. 

We show the influence of spatial resolution on the particle injection shares in Figure~\ref{fig:res_conv_inj_shares}. In order to display only statistically significant results, we only show injection shares after~$0.1\%$ of the total injected tracer particle population has undergone injection. We find that the injection shares have more difficulty converging, with no clear monotonic trend emerging upon varying~$d_e/\Delta x = 1.5 \to 8.0$. However, we find that~$\lambda_{\rm De}/\Delta x < 1$ causes a significant break in the results (panel~b).

\subsection{Particles per cell convergence study}

\begin{figure}
    \centering
    \includegraphics[width=\textwidth]{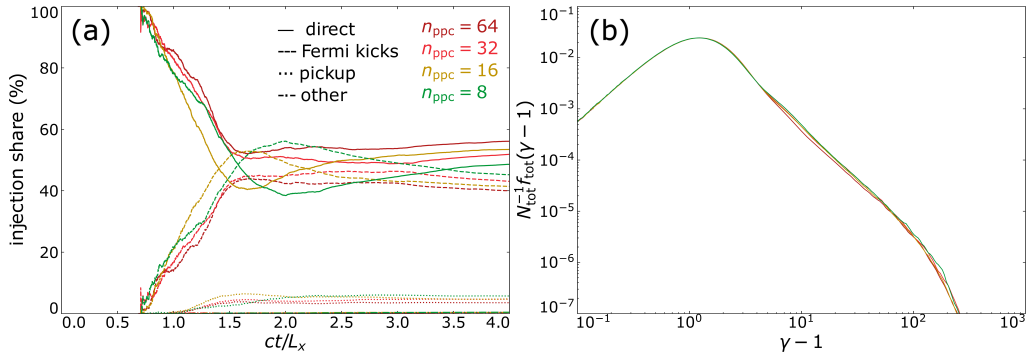}
    \caption{Panel~(a): Particle spectra for different~$n_{\rm ppc}$. Panel~(b): Injection shares for different~$n_{\rm ppc}$.}
    \label{fig:nppc_shares_spectra}
\end{figure}

In this study we strive to find a definite~$n_{\rm ppc}$ (number of particles per cell \textit{per species}) above which the results are insensitive and below which the results are sensitive. We find negligible variation for the injection energies and efficiencies, owing to well-converged spectra (Fig. \ref{fig:nppc_shares_spectra}b). However when checking injection shares, we find a significant difference between~$n_{\rm ppc} = 16$ and~$n_{\rm ppc} = 32$, especially for~$ct/L_x \in [1, 2]$ (Fig. \ref{fig:nppc_shares_spectra}a). Meanwhile, $n_{\rm ppc} = 32, 64$ yield similar injection shares for all mechanisms, remaining within~$5\%$ during the entire evolution. Therefore, we determine that~$n_{\rm ppc} = 32$ captures the results of greater~$n_{\rm ppc}$. Preserving the number of particles per Debye volume, this implies that we use~$n_{\rm ppc}^{(\text{3D})} = \frac{1}{2} \big( 2\, n_{\rm ppc}^{(\text{2D})} \big)^{2/3} = 8$ in 3D simulations.

\newpage
\section{Spectral fitting procedure} \label{sec:Appendix_procedure}

To measure the injection energies, we identically follow the fitting procedure described in \citet{French_2023} bar the following improvements. The first is that, rather than scanning over a range of~$\gamma_{\rm mono}$ values, we set~$\gamma_{\rm mono}$ to twice the energy of the spectral peak of the downstream particles. This simplifies the procedure and reduces the number of inputs without incurring much difference to the results. Second, instead of setting ``the" power-law index as the median of the collection of power-law indices that fall within a certain tolerance, we collect the median power-law index over the interval (denoted~$p_{\rm med}$) and plug in power-law indices~$p_j$ around it~$p_j \in [p_{\rm med} - \delta p, p_{\rm med} + \delta p]$ into the expression

$$ \mathcal{I}_j \equiv \int_{\gamma_1}^{\gamma_2} \bigg( \frac{A_j\gamma^{-p_j}}{f_{\rm ds}(\gamma)} - 1 \bigg) \,d\gamma, \ \ A_j \equiv \text{max}(f_{\rm ds}\gamma^{p_j}), $$
where~$\gamma_1$ and~$\gamma_2$ are the energies that bound the longest logarithmic power-law segment that resides within the power-law tolerance. The goal is to determine by brute force the~$p_j$ that yields the minimum~$\mathcal{I}_j$ in the set~$\{ \mathcal{I}_j \}$. Notice that~$A_j$ is defined to ensure that the fit intersects~$f_{\rm ds}$ exactly once. We choose~$\delta p = 0.5$ because that is sufficiently large to capture the best power-law index available. 

\begin{figure}
    \centering
    \includegraphics[width=\textwidth]{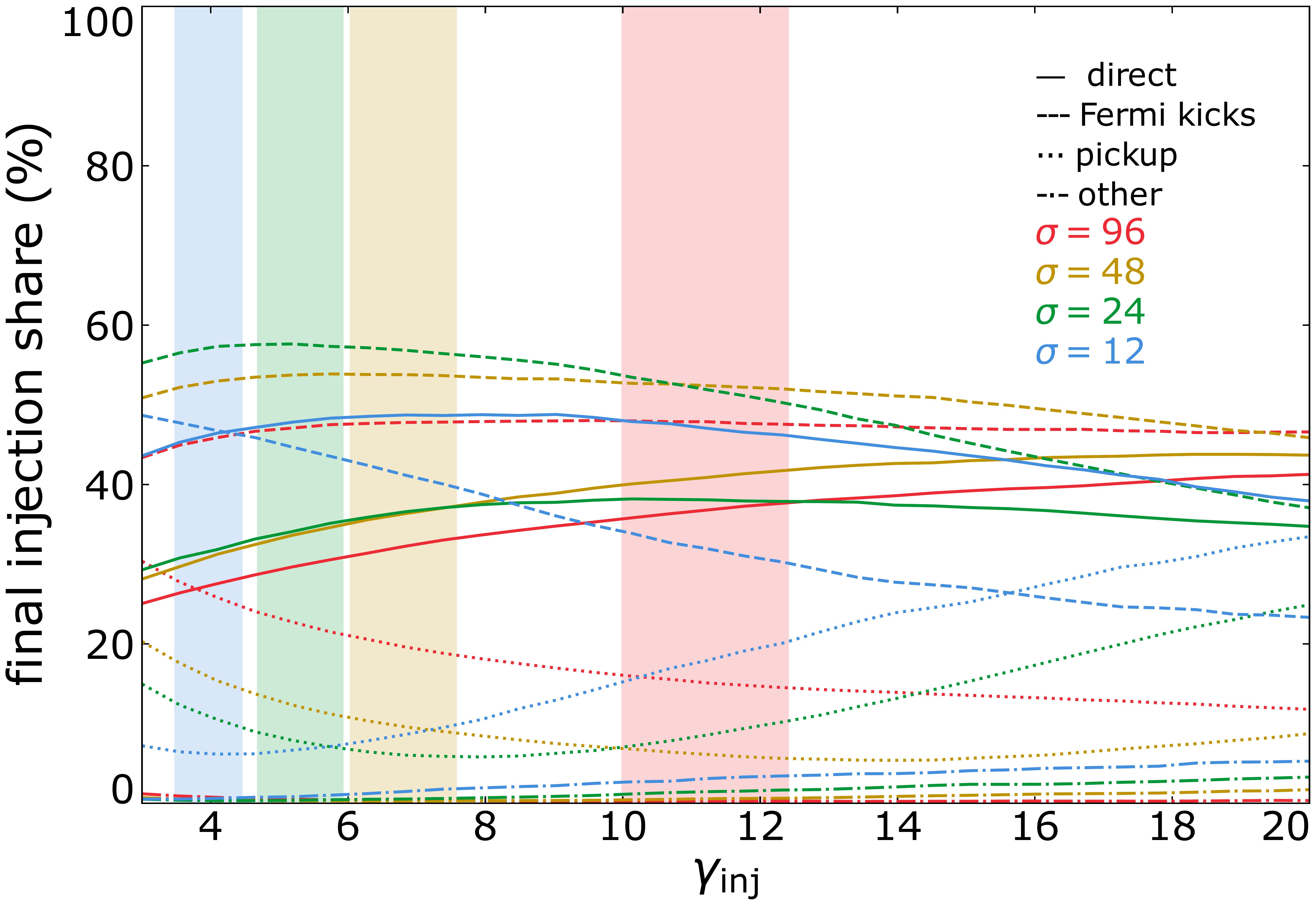}
    \caption{Injection shares at~$t = t_{\rm onset} + 3\, L_x/c$ as a function of~$\gamma_{\rm inj}$ for 2D simulations. The solid bars indicate the error about the measured injection energies. 
    }
    \label{fig:injection_shares_vs_ginj}
\end{figure}

Owing to these improvements to the fitting procedure, we are able to employ stricter power-law tolerances, $p_{\rm tol} \in [0.05, 0.20]$ and a smaller~$\alpha \equiv p - p_\gamma = 0.05$ (where~$p_\gamma \equiv -\,d \log{f_{\rm ds}(\gamma)}/d \log{\gamma}$ is the local power-law index) to define the injection energy. By contrast, in our previous work we used~$p_{\rm tol} \in [0.10, 0.30]$ and~$\alpha = 0.2$ \citep{French_2023}. The measurement errors of the power-law indices are within~$\simeq 0.05$, injection energies $\gamma_{\rm inj}$ within~$\simeq 10\%$, and high-energy cutoffs~$\gamma_c$ within~$\simeq 10\%$.

To show the insensitivity of the choice of~$\alpha$ on the final injection shares, we show the final injection shares against~$\gamma_{\rm inj}$ in Figure~\ref{fig:injection_shares_vs_ginj}. From the negligible variation of the final injection shares over each shaded bar, we see that regardless of whether~$\alpha$ is chosen to be~$0.05$ (as in this work) or~$0.2$ as in \citet{French_2023}, the injection shares are not significantly altered. 

\newpage

\bibliographystyle{jpp}
\bibliography{main}

\end{document}